\documentclass[final,3p,times]{elsarticle}
\usepackage[T1]{fontenc}
\usepackage{amssymb}

\usepackage{pdfpages}
\usepackage[colorinlistoftodos]{todonotes}
\usepackage{subfig}
\usepackage{colortbl}
\usepackage{subfig}
\usepackage{caption}
\usepackage{subcaption}
\usepackage{dblfloatfix}
\usepackage{svg}
\usepackage{amsmath}
\usepackage{float}
\usepackage{multicol}
\usepackage{comment}
\usepackage{lineno}
\usepackage{siunitx}  
\usepackage{lipsum}
\usepackage[section]{placeins}
\biboptions{sort&compress}
\journal{Journal}

\begin{document}

\begin{frontmatter}
\title{Sharp periodic Ge concentration modulations beyond the conduction band valley wavevector $k_0$ in nuclear spin-free Si quantum wells}
\author[1]{Ivo Rahlff}
\author[1]{Carsten Richter}
\author[1]{Martin Schmidbauer}
\author[1]{Maximilian Oezkent}
\author[1]{Thilo Remmele}
\author[2]{Michael Hanke}
\author[3]{Lars R. Schreiber}
\author[3]{Denny Dütz}
\author[3]{Sammy Umezawa}
\author[1]{Martin Albrecht}
\author[4]{Yann-Michel Niquet}
\author[5]{Tancredi Salamone}
\author[5]{Biel Martinez Diaz}
\author[1,6]{Thomas Schroeder}
\author[1]{Jens Martin}
\author[1]{and Kevin-P. Gradwohl\corref{cor1}}
\cortext[cor1]{kevin-peter.gradwohl@ikz-berlin.de}
\affiliation[1]{organization={Leibniz-Institut für Kristallzüchtung},
            state={Berlin},
            country={Germany}}
\affiliation[2]{organization={Paul-Drude-Institut für Festkörperelektronik, Leibniz-Institut im Forschungsverbund Berlin e.V.},
            state={Berlin},
            country={Germany}}
\affiliation[3]{organization={JARA-FIT Institute for Quantum Information, Forschungszentrum Jülich GmbH and RWTH Aachen University},
            city={Aachen},
            country={Germany}}
\affiliation[4]{organization={Université Grenoble Alpes, CEA, IRIG-MEM-L Sim},
            city={Grenoble},
            country={France}}            
\affiliation[5]{organization={Université Grenoble Alpes, CEA, Leti},
            city={Grenoble},
            country={France}}
\affiliation[6]{organization={Institut für Physik, Humboldt-Universität zu Berlin},
            city={Berlin},
            country={Germany}}

\begin{abstract}
Periodic Ge modulations within strained Si quantum wells in SiGe heterostructures offer a route to deterministically enhance conduction-band valley splitting in Si, a key requirement for scalable spin-qubit quantum computing. Efficient enhancement requires modulations in the order of the Si valley wavevector $k_0$ (9.7~nm$^{-1}$), corresponding to a period of 0.64~nm and near-monolayer growth control.

Using nuclear-spin-free molecular beam epitaxy with $^{28}$Si and $^{72}$Ge, we demonstrate Ge-modulated Si quantum wells with periods from 2.00 to 0.49~nm, including modulations at $k_0$ and $2k_0/3$. Synchrotron X-ray techniques and scanning transmission electron microscopy reveal laterally homogeneous Ge modulations over micrometer scales, with amplitudes up to 10~at-\% and gradients reaching 20~at-\%/nm. Two-bands $\mathbf{k}\cdot\mathbf{p}$ simulations suggest deterministic enhancement of valley splittings in steep trapezoidal $2k_0/3$ heterostructures, while the effect in $k_0$-type quantum wells is much weaker.
          
\end{abstract}

\end{frontmatter}
\begin{multicols}{2}

Electron spin qubits hosted in gate-defined quantum dots in planar \mbox{Si/SiGe} heterostructures are among the leading platforms for large-scale quantum computing \cite{Burkard2023, Schreiber2026}. These qubits can be controlled entirely electrically at high clock rates \cite{Philips22, defuentes2025, HRL2026}, while their manipulation, initialization, and readout fidelities have already surpassed the thresholds required for quantum error correction \cite{Noiri2022, Mills2022, Xue2022, Wu2025, HRL2026}. Further improvements in qubit coherence are expected through the systematic use of highly isotopically purified, nuclear-spin-free Si and Ge \cite{Struck2020, Stano22, cvitkovich2024coherence, HRL2026}. 
In addition, spin qubits in Si/SiGe feature an exceptionally small footprint. Combined with recent advances in conveyor-belt electron shuttling in Si/SiGe~\cite{desmet2024, Xue2024, Beer2026, Matsumoto2026}, these developments have stimulated proposals for scalable architectures comprising millions of qubits~\cite{Boter2022, kunne2024spinbus}, compatible with industrial silicon foundries~\cite{George2025, Muster2025Shuttling}.

\begin{figure*}[!t]
    \centering
    \includegraphics[width=1.0\linewidth]{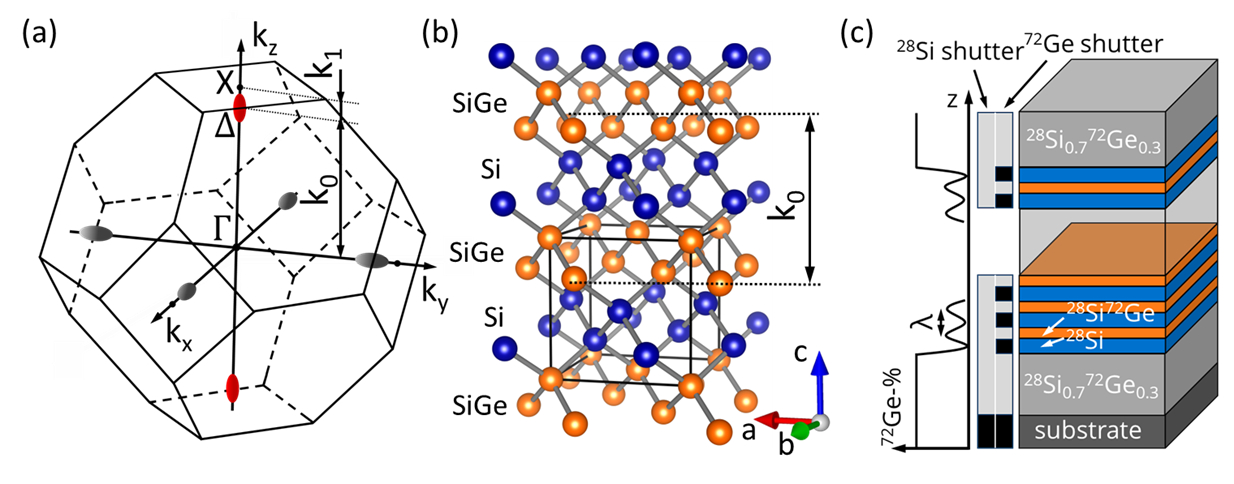}
    \caption{(a) Sketch of the Brillouin zone boundary of silicon with its conduction band minima depicted as ellipsoids and the relevant conduction band valley wavevectors $k_0$ and $k_1$. (b) Idealized atomic structure of a WQW with period length of a silicon lattice constant (0.543~nm) and the relative size of the conduction band wavevector $k_0$. (c) Schematic of WQW heterostructure by nuclear spin-free MBE together with the evaporation source shutter configuration and the respective $^{72}$Ge concentration profile.}
    \label{fig:CBM}
\end{figure*}

Despite this rapid progress, one major materials challenge remains: controlling the energy separation between the two low-lying valley states, commonly referred to as the valley splitting $E_\text{VS}$ \cite{cvitkovich2026valley, marcks2025valley, lima2024valley, Volmer2024}. Small valley splittings, $E_\text{VS} \lesssim \SI{50}{\micro\electronvolt}$, can compromise spin readout fidelity \cite{tagliaferri_impact_2018}, reduce coherence times \cite{hollmann_large_2020}, and limit the fidelity of coherent electron shuttling \cite{langrock_blueprint_2023, Volmer2026, losert_strategies_2024, david2024long}. 
Several approaches have been proposed to mitigate low $E_\text{VS}$ \cite{losert2023practical, stehouwer2025engineering}. However, many strategies primarily increase the average value of $E_\text{VS}$ without reliably eliminating local regions of statistically low $E_\text{VS}$ \cite{mcjunkin_valley_2021,wuetz_atomic_2022, klos_atomistic_2024}. One promising route toward a deterministic large valley splitting, i.e. a global $E_\text{VS}$ gap, is the periodic modulation of the Ge concentration, known as a  Wiggle-Quantum-Well (WQW) \cite{mcjunkin_sige_2022, feng2022enhanced,thayil2025theory,losert2023practical}.

In such WQWs, the Si quantum well (QW) is modulated by a periodic Ge concentration profile, which couples the degenerate conduction band minima by providing the corresponding wavevector. The most natural and reportedly most effective way to achieve large coupling and consequently large valley splitting is a modulation of the Ge concentration by 2$k_0$, with the conduction band valley wavevector $k_0$ within the Brillouin zone of Si shown in Fig. \ref{fig:CBM}a. This 2$k_0$ modulation equals 19.4 nm$^{-1}$ or a real space modulation of 0.32~nm (2.4 monolayers period length). While this approach is, in theory, the most effective, where even the smallest Ge modulations (below 1 at-\%) can achieve several meV of $E_\mathrm{vs}$, such material requires monolayer control in the related epitaxial processes and is consequently tremendously difficult to produce. Furthermore, one wants to introduce as little Ge as possible into the quantum well, to avoid unwanted alloy disorder and increased spin-orbit interaction \cite{woods2023spin}, which renders materials characterization of such WQWs extremely challenging. The shortest WQW periodicity so far reported is 1.8~nm (13 monolayers) \cite{feng2022enhanced,gradwohl2025enhanced}, motivated by a more complex inter-Brillouin zone coupling at a wavevector $k_1$, which requires additional strain engineering to be effective \cite{woods2024coupling}. However, WQWs exhibiting higher-order harmonics of the 2$k_0$ valley interference pattern (and longer period lengths) are also predicted to deterministically enhance valley splitting \cite{mcjunkin_sige_2022, CvitkovichStano26}. Consequently, the necessary epitaxial control to fabricate such WQW structures is slightly relaxed. 

In this work, we show epitaxy and materials characterization of such higher-order WQWs and simulate anticipated valley splittings. We report on WQWs with Ge modulations periods between 2.00~nm and 0.49~nm, using molecular beam epitaxy with nuclear spin-free source materials $^{28}$Si and $^{72}$Ge, with $^{29}$Si and $^{73}$Ge concentration below 100 ppm \cite{oezkent2026epitaxy}. 

The grown structures include second-order WQWs with wavevector $2k_0/2=k_0$ (0.64~nm period), and third-order $2k_0/3$ WQWs (0.97~nm period), with the latter promising the highest deterministic valley splitting enhancement in two bands $\mathbf{k}\cdot\mathbf{p}$ valley splitting simulations \cite{Salamone26}. The shortest WQW exhibits 0.49~nm period length, which is below the lattice constant of silicon as sketched in Fig.~\ref{fig:CBM}b, demonstrating epitaxial control below two monolayers.

\section*{Results}
\label{sec:results}

A series of WQWs were grown using low-temperature molecular beam epitaxy at 200°C growth temperature to suppress Ge segregation. The Ge modulations are achieved by using constant Si flux, while modulating the Ge shutter, as shown in Fig.~\ref{fig:CBM}c and elucidated in more detail in subsection \textit{MBE Growth}. The samples were grown with varying Ge concentration oscillation periodicities $\lambda_{\text{WQW}}$ in the Si quantum well (Fig.~\ref{fig:CBM}c). The list of WQW grown samples with measured Ge concentration modulation, and the used methods to detect the modulation, can be seen in the Table~\ref{tab:samples}.

\begin{table}[H]
\caption{Overview of the investigated WQW samples in this work. Extracted WQW periods $\lambda$ from GIXD measurements (cf. subsection \textit{Grazing Incidence Crystal Truncation Rod Analysis}), as well as their wavevector $q$, and with which x-ray methods could confirm the modulations. More detailed information about the epitaxial layer structure, i.e., target layer thicknesses and composition, are given in the supplementary information 1 (SI1).}
\begin{tabular}{|c|c|c|l|} 
\rowcolor[HTML]{FFCE93}
 \hline
 WQW & $\lambda$ (nm) & $q$ (nm$^{-1}$) & confirmed by \\  
 \hline
 A & 2.00 & 3.1  & XRR, GIXD  \\ 
 \hline
 B & 1.02 & 6.2 ($\approx 2k_0$/3)  & XRR, GIXD  \\
 \hline
 C & 0.88 & 7.1   & XRR, GIXD  \\
 \hline
 D & 0.62 & 10.1 ($\approx k_0$)  & GIXD  \\
 \hline
 E & 0.49 & 12.8  & GIXD  \\
 \hline
\end{tabular}
\label{tab:samples}
\end{table}

In-house X-ray specular reflectivity (XRR) measurements revealed a lower limit of detection of approximately 0.8~nm for the WQW period (see supplementary information 2 - SI2). Therefore, the use of high-brilliance synchrotron radiation was necessary for further investigations. This ultimately enabled us to detect a WQW period of 0.49 nm, i.e., a length that is smaller than the dimension of a crystal unit cell of 0.543~nm consisting of four monolayers. Grazing-incidence X-ray diffraction (GIXD) has proven to be more sensitive technique for WQWs with short period lengths than XRR, while in the latter, precise Ge concentration profiles could be extracted. Supplementary scanning transmission electron microscopy (STEM) results for one of the samples confirm the periodic layered structure of the investigated WQW and show a Ge concentration profile that agrees well with the profile extracted from the XRR simulation. This work will focus mainly on X-ray methods, due to their high sensitivity for WQW periodicity, fast feedback loop, and non-destructive nature. Consequently, these results will be decisive in the race to develop the next-generation spin-qubit material based on Si-QWs or, more precisely, WQWs.

\subsection*{X-Ray Specular Reflectivity}

\begin{figure*}[]
    \centering
    \includegraphics[width=0.9\linewidth]{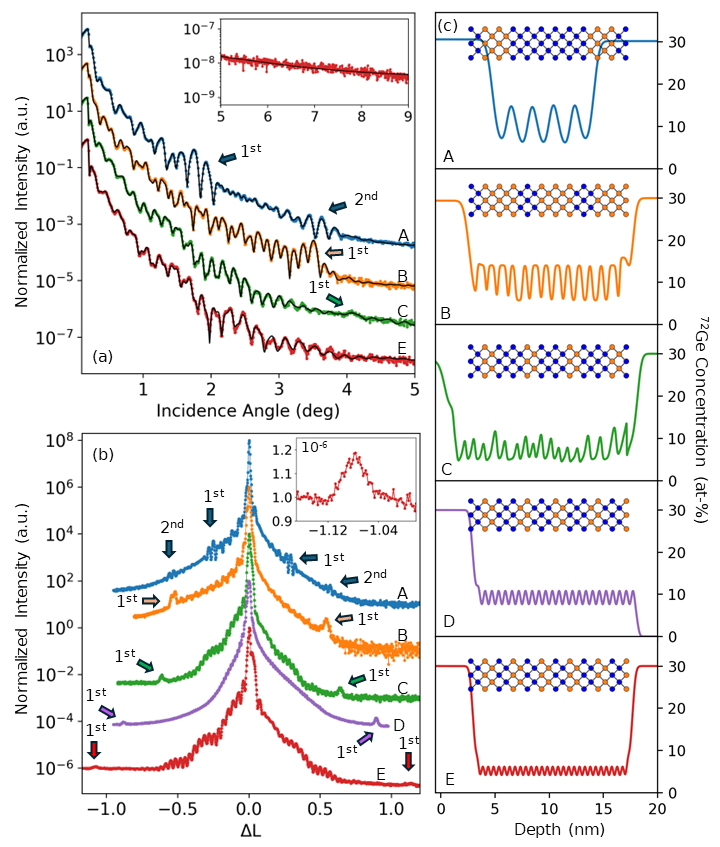}
    \caption{(a) X-ray reflectivity curves for samples A, B, C, and E (colored symbols) along with corresponding simulations (black lines) measured at an x-ray wavelength of $\lambda_{\text{xray}} = 1.2398$~$\AA$. The curves are normalized to their respective maxima and shifted on the y-axis for better comparability. The first- and second-order superlattice peaks of the WQWs are indicated by arrows. The inset shows corresponding measured data for sample E in an extended angular range that includes the predicted superlattice peak at approximately \SI{6.05}{\degree}. (b) GIXD CTR scans of the investigated WQW samples, measured along the vertical L-direction. For samples A, B and D the intensity in the vicinity of the Si 111 Bragg reflection ($\Delta L = 0$) is shown, whereas for samples C and E the intensity in the vicinity of the Si 202 Bragg reflection ($\Delta L = 0$) is shown. WQW superlattice peaks (of $1^{st}$ and $2^{nd}$ order) are observed for all samples . The experimental curves are normalized to its corresponding maximum, and shifted on the y-axis for better comparison. The inset shows the left WQW peak of sample E in more detail (linear scale). (c) Ge concentration profiles in the WQWs extracted from the X-ray simulations of samples A-C. The profiles are calibrated to correspond to an absolute concentration of Si$_{0.7}$Ge$_{0.3}$ in the lower barrier. The profiles of sample D and E are deduced from the WQW peak position and intensity from the corresponding CTR curves. The corresponding insets represent the approximated atomic layer structure of the samples, similar to Fig. \ref{fig:CBM} (b). More detailed information on the simulations and profile deduction are provided in SI3. }
    \label{fig:XRR_all_samples}
    
\end{figure*}

XRR is sensitive to the vertical electron density profile and is therefore frequently used to determine layer thicknesses, chemical composition, and interfacial widths (e.g. roughness) between adjacent layers. In particular, we have used this well-established technique for our samples in order to obtain detailed information about the mean WQW periodicity $\lambda_{\text{WQW}}$, the individual thicknesses for Si and Ge$_x$Si$_{1-x}$ of the WQW, and the corresponding Ge compositions $x$.

Fig.~\ref{fig:XRR_all_samples}a displays XRR curves of samples A-C and sample E as a function of the glancing angle of incidence $\theta$ measured at an X-ray wavelength of $\lambda_{\text{xray}} = 1.2398$~$\AA$. They exhibit several prominent features, like their maximum intensity at the critical angle of total external reflection (here at approx. $\theta = \theta_C = $ \SI{0.18}{\degree}), interference fringes (\textit{Kiessig fringes}) and a strong decay of the intensity with increasing incidence angle. For \mbox{samples A}, B and C evidence for the periodicity of the WQWs is reflected as characteristic superlattice peaks, which are highlighted by the arrows in the graph. 

Neglecting X-ray refraction effects occurring at small angles of incidence, we can directly use the superlattice peak positions $\theta_{SL}$ to estimate the WQW period $\lambda_{\text{WQW}}$. Periods of (1.96$\pm$0.04)~\si{\nm}, (1.01$\pm$0.03)~\si{\nm} and (0.87$\pm$0.03)~\si{\nm} for \mbox{sample A}, \mbox{sample B} and \mbox{sample C}, respectively, are extracted. 

More accurate and extended information on individual layer thicknesses, corresponding interfacial roughnesses and Ge concentration profiles has been extracted from simulations using dynamical diffraction theory. In SI4, a detailed overview of the models, the simulation approach, and the accuracy and reliability of the fitting procedure is given. 

The results for sample A-C are presented in Fig.~\ref{fig:XRR_all_samples}a as solid black lines with excellent agreement to the experimental curves, and the respective concentration profiles in Fig.~\ref{fig:XRR_all_samples}c. The average WQW parameters such as Si and SiGe layer thickness, as well as the concentration slopes, are summarized in Table~\ref{tab:stats_XRRSim}. The Ge concentration slopes range somewhere from 10~at-\% to 20~at-\%, with sample B having the steepest concentration gradients.


\begin{table*}[]
\caption{Extracted WQW layer parameters from the XRR simulations illustrated in Fig. \ref{fig:XRR_all_samples}. The layer thicknesses were extracted directly from the output parameters of the corresponding simulation. The Ge concentration slopes were derived from the Ge concentration profiles given in Fig. \ref{fig:XRR_all_samples} (c). Furthermore, the Ge concentration slopes, derived from the STEM-EDS measurement, are given in the table (blue cells). Detailed information about the simulation parameters and slope extraction is given in SI1 and SI5.}
\label{tab:stats_XRRSim}
\centering
\begin{tabular}{|c|c|c|cl|cl|}
\hline
\rowcolor[HTML]{FFCE93} 
Sample & \begin{tabular}[c]{@{}c@{}}Si layer \\ thickness (nm)\end{tabular} & \begin{tabular}[c]{@{}c@{}}SiGe layer \\ thickness (nm)\end{tabular} & \multicolumn{2}{c|}{\cellcolor[HTML]{FFCE93}\begin{tabular}[c]{@{}c@{}}Si-SiGe concentration \\ slope (at-\%/nm)\end{tabular}} & \multicolumn{2}{c|}{\cellcolor[HTML]{FFCE93}\begin{tabular}[c]{@{}c@{}}SiGe-Si concentration \\ slope (at-\%/nm)\end{tabular}} \\ \hline
A      & 1.45 $\pm$ 0.09                                                    & 0.53 $\pm$ 0.04                                                      & \multicolumn{1}{c|}{8.5 $\pm$ 0.1}                           & \cellcolor[HTML]{DAE8FC}12.8 $\pm$ 3.5                           & \multicolumn{1}{c|}{9.7 $\pm$ 0.2}                           & \cellcolor[HTML]{DAE8FC}8.6 $\pm$ 1.3                           \\ \hline
B      & 0.40 $\pm$ 0.14                                                    & 0.62 $\pm$ 0.14                                                      & \multicolumn{2}{c|}{20.5 $\pm$ 0.6}                                                                                            & \multicolumn{2}{c|}{20.7 $\pm$ 0.5}                                                                                            \\ \hline
C      & 0.69 $\pm$ 0.19                                                    & 0.22 $\pm$ 0.11                                                      & \multicolumn{2}{c|}{15.4 $\pm$ 1.9}                                                                                                & \multicolumn{2}{c|}{9.2 $\pm$ 1.7}                                                                                             \\ \hline
\end{tabular}
\end{table*}

\label{subsec:XRRLimit}

With decreasing WQW periodicity $\lambda_{\text{WQW}}$, the positions of the (first-order) superlattice peaks shift to higher angles, while at the same time their intensities decrease dramatically, challenging materials characterization. In case of sample E, with a nominal WQW period of 0.49 nm, no first order WQW superlattice peak \mbox{- which is expected to appear at about \SI{6.05}{\degree} -} can be observed (see inset of Fig.~\ref{fig:XRR_all_samples}a). There are various reasons for this decrease in intensity, all of which we list and discuss in detail in SI3. To summarize the results of these considerations, we find that the intensity of the superlattice peak decreases as the third power of the WQW period, such that very short periods no longer exhibit detectable intensity and fade into the scattering background.

\begin{figure*}[!t]
    \centering
    \includegraphics[width=1.0\linewidth]{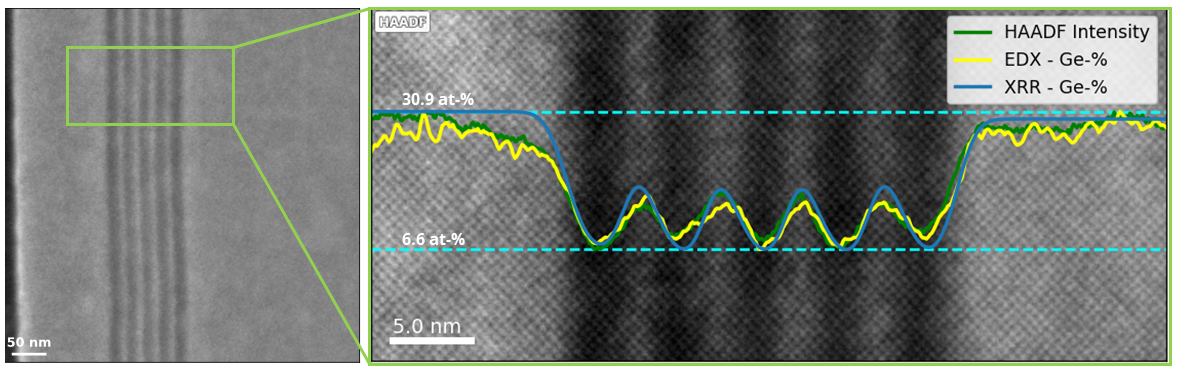}
    \caption{Overview STEM graph of sample A (2.00nm-WQW). The close-up shows the HAADF graph, with its intensity profile (averaged over the area between the cyan lines) overlayed in the graph. Furthermore the Ge concentration profile derived from the STEM-EDX measurement (of the same area) and the Ge concentration profile derived from the XRR simulation of sample A is shown. For comparison, all three profiles were scaled and adjusted to each other, whereby the XRR profile is given in at-\%.}
    \label{fig:STEM}
\end{figure*}

\subsection*{Scanning Transmission Electron Microscopy}

To support our findings, one of the investigated samples was characterized by means of STEM. In that regard, a (1~1~0) cross-section of sample A was investigated by high angular annular dark field (HAADF)-STEM and STEM-based energy dispersive x-ray spectroscopy (EDX), as shown in Fig.~\ref{fig:STEM}. The image reveals a WQW with pronounced Ge modulations and a periodicity of (2.04 $\pm$ 0.18)~\si{\nm}. The bending of the interfaces in the STEM image is likely an artifact due to warping of the sample lamella. The close-up in Fig.~\ref{fig:STEM} shows an atomic resolution zoom-in, giving further insight in the interface morphology and interface sharpness.\newline
\newline
The superimposed  profiles in the close-up are the HAADF intensity and the EDX Ge concentration (linear correction for thickness contrast) together with the Ge concentration profile derived from the XRR simulation. From the EDX profile slopes of the Ge concentration of about (12.8 $\pm$ 3.5)~at-\%/nm and (8.6 $\pm$ 1.3)~at-\%/nm can be derived for the Si-SiGe and SiGe-Si interface, respectively (cf. SI6). The data agrees to a high degree, validating our XRR fitting procedure, and generally the X-ray based WQW characterization approach.

\subsection*{Grazing Incidence Crystal Truncation Rod Analysis}
\label{subsec:GIXD}

XRR fails for WQW periods shorter than approximately 0.87 nm due to an insufficient signal-to-noise ratio at large angles of incidence, where the penetration depth is far greater than the depth of the buried WQWs. We can overcome this limitation by applying crystal truncation rod (CTR) analysis performed at fixed glancing angle of incidence. Selecting an angle of incidence close to the critical angle of total external reflection selectively enhances the scattering caused by the WQWs, while the undesired scattering signal from the underlying SiGe barrier and the SiGe substrate is strongly suppressed. For samples A, B, and D the CTR in the vicinity of the 111 Bragg reflection was investigated, whereas the 202 Bragg reflection was chosen for samples C and E. This choice is particularly advantageous because the 110, 112, 201, and 203 Bragg reflections are forbidden in the diamond crystal structure, resulting in low background intensity, especially for samples with a WQW period approaching the size of a Si unit cell. These WQWs should exhibit weak superlattice peaks close to those forbidden reflections.


The GIXD CTR measurements of all samples are summarized in Fig. \ref{fig:XRR_all_samples}b. In addition to the sharp substrate Bragg peak at $\Delta L = 0$, all curves exhibit intensity oscillations originating from the top Si$_{0.7}$Ge$_{0.3}$ barrier (if present) and the WQWs. In particular, superlattice Bragg reflections (marked by corresponding arrows) are observed, from whose positions the corresponding period $\lambda_{\text{WQW}}$ can be deduced. Due to the greatly improved signal-to-noise ratio, a corresponding, albeit very weak, scattering signal is also present for both the 0.62~nm sample and the 0.49~nm sample.


Similar to the XRR measurements shown in Fig.~\ref{fig:XRR_all_samples}a, the interference fringes in the GIXD rapidly drop below the noise level starting at a $\Delta$L of about $\pm 0.5$, while the WQW Bragg peaks rise above the background level at higher magnitutes of $\Delta$L. Even for samples D and E with a period of $\lambda_{\text{WQW}}$ = 0.62~nm and $\lambda_{\text{WQW}}$ = 0.49~nm, respectively, - and for which our XRR analysis fails - first order superlattice peaks can be clearly identified (see inset of Fig. \ref{fig:XRR_all_samples}b for sample E). Our data therefore provide clear evidence of strongly correlated Ge composition modulations that constructively interfere on the scale of the X-ray photon coherence length, confirming laterally homogeneous WQW modulations of at least several \si{\micro m}.

The WQW periodicity can be easily extracted from the L-position of the diffraction peaks: $\lambda_{\text{WQW}} = \frac{2a_{\text{sub}}}{\Delta\text{L}_{WQW}}$, where $a_{\text{sub}}$ is the lattice parameter of the Si$_{0.7}$Ge$_{0.3}$ virtual substrate and $\Delta\text{L}_{\text{WQW}}$ the distance between opposite first order superlattice peaks. Oscillation periods of (2.00 $\pm$ 0.05)~\si{\nm}, (1.02 $\pm$ 0.03)~\si{\nm}, (0.88 $\pm$ 0.02)~\si{\nm}, (0.62 $\pm$ 0.01)~\si{\nm}, and (0.49 $\pm$ 0.02)~\si{\nm}, for A, B, C, D, and E, respectively, are extracted. For samples A, B, and C  excellent agreement with the results obtained via XRR is achieved. Additionally, the Ge amplitude of the WQWs can be estimated from the signal to noise ratio of the main diffraction peak relative to the WQW satellite peaks, which suggest a Ge amplitude of 8~at-\% and 5~at-\% for sample D and E (more details in SI7), respectively. This data was used to estimate the Ge concentration profiles of these samples in Fig.~\ref{fig:XRR_all_samples}c.

\subsection*{Valley-Splitting Simulations}

The measurement of valley splittings in Si/SiGe quantum dots for spin qubits is experimentally complex, requiring full fabrication of quantum chips (including implantation, thermal treatments, gate stack deposition...) and cryo-electronic characterization \cite{Volmer2024, Volmer2026_complex}. Nevertheless, all of that should not change the WQW profiles. Here, we demonstrate that such WQWs can be produced by epitaxy in the first place, and we use their structural data as input for valley splitting simulations that explore the anticipated valley splitting enhancement.

We calculate the valley splittings $E_\mathrm{vs}$ in these heterostructures using a two-bands $\mathbf{k}\cdot\mathbf{p}$ (2KP) model \cite{Salamone26} as well as an atomistic tight-binding (TB) model \cite{Niquet09}. The TB model is expected to provide the most accurate description of valley physics while the 2KP model is much more efficient and can address larger structures. We consider quantum dots confined vertically by the Ge concentration profiles of Fig. \ref{fig:XRR_all_samples}c and a constant vertical electric field $E_z$, and laterally by a harmonic potential characterized by the in-plane dot radius $r_\parallel$ (ranging from 5 to 20 nm). We account for alloy disorder by randomly distributing the Ge atoms in each monolayer. We collect statistics [median valley splitting $\overline{E_\mathrm{vs}}$ and inter-quartile range (IQR)] over 256 realizations of this disorder. Details about the models and methodology can be found in Ref. \cite{Salamone26}.

In order to assess the effects of the Ge modulations, we compare each WQW structure to a plain quantum well with uniform Ge concentration (the average Ge content $c_\mathrm{avg}$ in the WQW), for different dot radii $r_\parallel$. In such a plain quantum well, the valley splitting is opened by alloy disorder, which is varying fast enough to exhibit significant Fourier components around the inter-valley wavevector $q=2k_0$ \cite{losert2023practical}. The median valley splitting $\overline{E_\mathrm{vs}}$ and the IQR are thus expected to decrease with increasing $r_\parallel$ as the alloy disorder gets averaged out on the scale of the dot. They remain, however, comparable, because there is always a fraction of quantum dots with small valley splittings. The valley splitting is deterministically enhanced by the WQW only if the median is much larger than in the plain quantum well ($\overline{E_\mathrm{vs}}\gg\mathrm{IQR}$) and is robust with respect to dot size.

\begin{figure*}[]
\includegraphics[width=\linewidth]{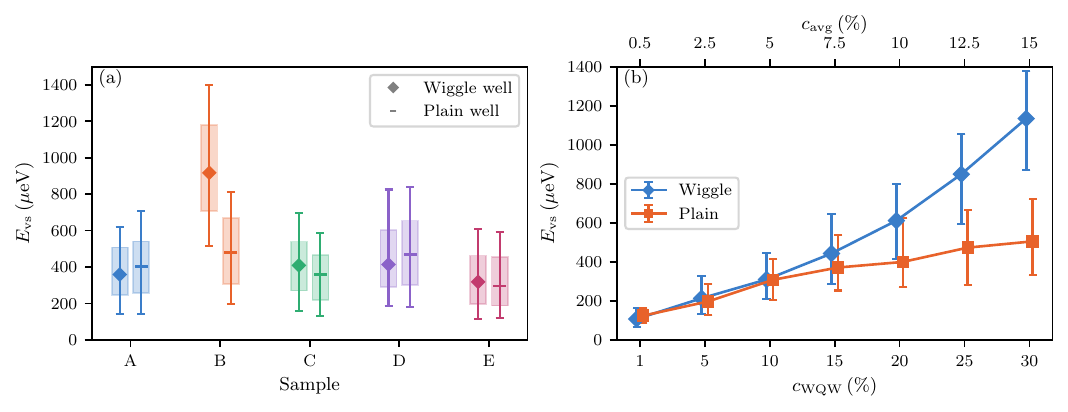}
\caption{(a) Box plots (median, inter-quartile and inter-decile ranges) of the valley splitting distributions computed with the 2KP model in samples A-E and in corresponding ``plain'' quantum wells with a uniform concentration of Ge (the average Ge content in the wiggle well). The radius of the dots is $r_\parallel=15$\,nm and the vertical electric field is $E_z=5$\,mV/nm, except for sample B ($E_z=-5$\,mV/nm). (b) Valley splitting as a function of the amplitude $c_\mathrm{WQW}$ of the wiggles in ideal WQWs with wave length $\lambda_{WQW}=2\pi/k_0$ and in plain wells with same average Ge content $c_\mathrm{avg}$, for dots with radius $r_\parallel=15$ nm. The median splittings are plotted with error bars giving the inter-quartile range. The vertical electric field is $E_z=5$\,mV/nm.} 
\label{figuresVS}
\end{figure*}


We show the calculated box plots (median, inter-quartile and inter-decile ranges of the valley splittings) of each structure in Fig. \ref{figuresVS}a, for a dot radius $r_\parallel=15$\,nm (2KP calculations). This radius is comparable to the estimated size \cite{Salamone26} of the dots of Ref. \cite{Philips22}. Only sample B (with steepest concentration gradients) allows for a clear enhancement of valley splittings, yet for negative electric fields (electron localized toward the bottom interface). The median valley splitting $\overline{E_\mathrm{vs}}$ of this sample indeed reaches $\approx 900\,\mu$eV, and is weakly dependent on dot size, whereas the IQR decreases with increasing $r_\parallel$ (see SI8). These conclusions are supported by TB calculations on smaller dots ($r_\parallel=8$\,nm), which confirm that only sample B reaches valley splittings near the meV range.

The Fourier transform of the Ge concentration profile of sample B actually exhibits large Fourier components around wavevector $q=2k_0$. This results from the steeper (trapezoidal more than sinus-like) modulations of the Ge concentration profile at wavevector $k\approx 2k_0/3\approx 6.2$\,nm$^{-1}$. As a consequence, this profile has strong harmonics at $q=nk$ (whose amplitude would decrease as $1/n$ for square modulations), so that the third harmonic at $q\approx 2k_0$ can still give rise to large valley splittings \cite{CvitkovichStano26}. We discuss in SI8 the dependence of the valley splitting in sample B on the vertical electric field, and how a careful engineering of the period, amplitude and shape of such ``trapezoidal'' WQWs can provide robust valley splittings with state-of-the-art growth techniques.

Sample D with wiggles at wave length $\lambda_{\text{WQW}}\sim 2\pi/k_0=0.64$\,nm is expected to show sizable a increase of the valley splittings (as a subharmonic of the $2k_0$ resonance) \cite{feng2022enhanced}. However, the effects of the small modulations of the Ge concentration are washed out by alloy disorder. Only sizable modulations ($>15\%$) of the Ge concentration at wavevector $k=k_0$ do enhance the valley splittings with respect to a plain quantum well in large enough dots (see Fig. \ref{figuresVS}b and SI8).


\section*{Discussion}

We demonstrate the feasibility of epitaxial growth of ultra-low-period nuclear spin-free Ge modulations in Si quantum wells. In particular, we show WQWs with periodicities down to 0.49~nm, including WQWs with wavevectors close to $k_0$ (0.64~nm period), and 2$k_0$/3 (0.97~nm period). We validate our claims by advanced non-destructive materials characterization tailored to analyze WQW modulations using X-ray techniques. The validity of X-ray results are confirmed by STEM investigations. 

We calculate the anticipated deterministic enhancement of the valley splittings in the grown structures with 2KP and TB simulations. On the one hand, $k_0$-WQWs appear to be a rather ineffective approach, exhibiting deterministic enhancement only above a Ge modulation of 15 at-\%. On the other hand, the simulations suggest a completely new way to take advantage of the steep SiGe slopes of 20~\%/nm, by designing efficient ``trapezoidal'' WQWs at a wavevector of 2$k_0$/3 (or more generally $2k_0/p$, $p$ odd integer), as further discussed in SI8. Consequently, among the grown structures, sample B, a trapezoidal 2$k_0$/3-WQW with steep SiGe slopes, turned out to be most promising, with valley splitting enhancements up to around 500~$\mu$eV.




Furthermore, we would particularly like to highlight the WQW with a period of $\lambda_{\text{WQW}}$ = 0.49~nm, which is below the lattice parameter of Si and Ge or four monolayers (Fig.~\ref{fig:CBM}b), and signifies that the thicknesses of the individual Si and SiGe layers are smaller than two monolayers. 

This underlines the high epitaxial control that is possible in low-temperature MBE equipment, and opens the door for a range of new Si-based applications, be it electronics, spintronics, or quantum applications that require single conduction bands in silicon, but also applications building upon symmetry breaking of the Si lattice. Since such SiGe superlattice periodicities of the order of the lattice constant also effectively break crystal symmetry, the heterostructure essentially exhibits different desirable properties, such as transition to direct bandgap semiconductors, Brillouin zone folding, but also particular phonon properties (e.g. mirrors) for advanced phononics. 

Ultimately, our results also push the boundary toward epitaxy of lower periodicity WQWs, once thought impractical, making the realization of even 2$k_0$ (0.32~nm) WQWs considerably more likely.

\section*{Online Methods}
\label{sec:methods}

\subsection*{MBE Growth}
\label{subsec:MBE_growth}
The epitaxial growth of the WQW samples is performed in a MBE system with an electron beam evaporation source and resistive heater source for Si and Ge, respectively. The MBE chamber has a base pressure in the order of \SI{e-11}{\milli \bar} and a stable growth pressure in the range of \SI{e-10}{\milli \bar}. All sample substrates were sewn to $25$ × $25$ \si{\milli \m \squared} from one batch of 12" Si (0 0 1) wafers. The wafers have a multilayered CVD grown structure and terminate with a Si$_{0.7}$Ge$_{0.3}$ fully relaxed layer. The terminated layer is chemical-mechanical polished, and epi ready.  A deviation from (0 0 1) of $<$ 0.1$^{\circ}$ was confirmed by X-ray diffraction (XRD) measurements. Prior to the growth, the substrates underwent a wet-chemical cleaning. After the wet-chemical cleaning, the substrate is transferred, with a total exposure to ambient air of $< 15$~s, into a load lock. The wet chemical cleaning process is intended to remove residual metals, oxides and carbon with leaving an oxide or hydrogen passivated surface. After the transfer of the substrate to the MBE chamber, the substrate is annealed at 650~°C in order to desorb hydrogen and residual oxygen from the substrate surface. The bottom barriers of the WQW samples were grown at a temperature of \SI{400}{\degree C}. During the growth of the QW the temperature was constant at \SI{200}{\degree C}. For the growth of the top barrier layers, the temperature was slightly increased ($<$\SI{230}{\degree C}). The growth rate was chosen to be approximately \SI{0.02}{\nm \per \sec}. The Ge modulations within the QW were achieved by varying the shutter state (open/close) of the Ge source periodically. We can define the Ge shutter ratio as $r_{\text{Ge}}=t_{\text{Ge}}/t_{\text{WQW}}$. Here $t_{\text{Ge}}$ is the Ge shutter time in its open state and $t_{\text{WQW}}$ is total growth time for one WQW period.

\subsection*{Synchrotron X-Ray Investigations}
The X-ray investigations were conducted using brilliant synchrotron radiation during two different beamtimes at PHARAO beamline at BESSY (Berlin, Germany) and at station BM 25 at ESRF (Grenoble, France) at photon energies of 10 keV ($\lambda_{\text{xray}} = 1.2398 \AA$) and 18 keV ($\lambda_{\text{xray}} = 0.6888 \AA$), respectively. 

\subsection*{X-Ray Reflectivity (XRR)}

For XRR incident slits of about 0.1 mm were used to guaranty that the x-ray beam is solely located on the sample surface and does not lead to additional scattering background from e.g. the sample holder. A two-dimensional detector (EIGER2 1M,  DECTRIS) with a pixel size of 75 $\mu m$ x 75 $\mu m$  was placed typically 1000 mm behind the sample ensuring sufficient angular resolution of order of 0.01°. The area detector was used to probe reciprocal space in 3D, however, a virtual exit slit could be defined on the detector to exclusively record the specularly reflected intensity and thus to perform line scans in reciprocal space. Additional XRR measurements were performed using a commercial, laboratory-based 9-kW SmartLab system (Rigaku) with Cu-K$\alpha 1$ radiation ($\lambda_{\text{xray}} = 1.5406 \AA$) and a corresponding beam collimation of approximately 0.008°, albeit with significantly lower counting statistics at high Q-values.

\subsection*{Grazing Incidence Crystal Truncation Rod (CTR) Scans}
The CTRs measurements were performed at a fixed angle of incidence slightly above the critical angle of total external reflection to ensure maximum sensitivity to the WQWs and to suppress the contributions of the virtual substrate underlying Si$_{0.7}$Ge$_{0.3}$. Here, again, the 2D detector was used to obtain the full 3D intensity distribution in the vicinity of selected reciprocal lattice points, from which “vertical” CTR scans were extracted.

\subsection*{STEM Scan}
STEM with EDX was done on a probe corrected transmission electron microscope equipped with an EDX system. The microscope was operated at 80 kV acceleration voltage to prevent electron beam induced damage of the WQW which occurred previously at 300 kV voltage. In addition the beam current was reduced to about 40 pA. The semi convergence angle of the probe was 21.4 mrad and the collection range of the high angular dark field detector was 41 - 200 mrad. The EDX data was recorded using a dispersion of 2 eV, a dwell time of 10 µs, and an integration of 40 frames with drift compensation. The thickness of the sample in [100] orientation was estimated by electron energy loss spectroscopy to be about 50 nm.

\section*{Acknowledgement}
\label{sec:methods}

We acknowledge the funding by the Deutsche Forschungsgemeinschaft (DFG, German Research Foundation) within the research project "Valley Splitting Engineering in Nuclear Spin-free SiGe for Silicon-Qubits" with the grant Nr. 554676597. We thank the ESRF, Grenoble for providing beamtime at SpLine BM25 (project HC-6506). The authors thank Juan Rubio-Zuazo (BM25, ESRF, Grenoble) for technical support. We are also grateful for beamtime granted at BESSY II, Helmholtz-Zentrum Berlin (through project 242-12757ST). YMN, TS, and BMD acknowledge support from the ''France 2030'' program (PEPR PRESQUILE-ANR-22-PETQ-0002) and from the Horizon Europe Framework Program (Grant Agreement No. 101174557 QLSI2).

\FloatBarrier
\newpage
\bibliographystyle{ieeetr} 

\bibliography{Library_XRR_Paper}
\end{multicols}
\FloatBarrier
\newpage
\clearpage


\section*{Supplementary material}

# Supplementary Information: Sharp periodic Ge concentration modulations beyond the conduction band valley wavevector $k_0$ in nuclear spin-free Si quantum wells

Ivo Rahlff[a], Carsten Richter[a], Martin Schmidbauer[a], Maximilian Oezkent[a], Thilo Remmele[a], Michael Hanke[b], Lars R. Schreiber[c], Denny Dütz[c], Sammy Umezawa[c], Martin Albrecht[a], Yann-Michel Niquet[d], Tancredi Salamone[e], Biel Martinez Diaz[e], Thomas Schroeder[a,f], Jens Martin[a], and Kevin-P. Gradwohl[a,*]

[a]*Leibniz-Institut für Kristallzüchtung, Berlin, Germany*
[b]*Paul-Drude-Institut für Festkörperelektronik, Leibniz–Institut im Forschungsverbund Berlin e.V., Berlin, Germany*
[c]*JARA-FIT Institute for Quantum Information, Forschungszentrum Jülich GmbH and RWTH Aachen University, Aachen, Germany*
[d]*Université Grenoble Alpes, CEA, IRIG-MEM-L Sim, Grenoble, France*
[e]*Université Grenoble Alpes, CEA, Leti, Grenoble, France*
[f]*Institut für Physik, Humboldt-Universität zu Berlin, Berlin, Germany*

## SI1: Target Sample Structures

In Tab. 1 the target sample structures are summarized in a tabular format. The layer thicknesses are indirectly given by the growth time $t$ and the calibrated/estimated growth rates $v_g$. The (expected) Ge concentration is given by the ratio of the growth rates $v_{g,\mathrm{Si}}$ and $v_{g,\mathrm{Ge}}$, whereby segregation or other effects are neglected. Sample A-C and E have the same sample structure, i.e., a $Si_{1-x}Ge_x$ bottom barrier, the WQW and a $Si_{1-x}Ge_x$ top barrier. Sample D has a slightly different sample structure. Sample D consists of a $Si_{1-x}Ge_x$ bottom barrier, a Si interlayer and the WQW.

Table 1: Target sample structures

| Sample ID | Layer | Epilayer | Periods | Thickness (nm) | Ge-concentration (atm-%) |
|---|---|---|---|---|---|
| A | Bottom Barrier | $Si_{1-x}Ge_x$ | 1 | 20 | 30 |
| | WQW | Si | 4 | 1.2 | 0 |
| | | $Si_{1-x}Ge_x$ | | 0.6 | 30 |
| | | Si | 1 | 1.2 | 0 |
| | Top Barrier | $Si_{1-x}Ge_x$ | 1 | 12 | 30 |
| | | | | | |
| B | Bottom Barrier | $Si_{1-x}Ge_x$ | 1 | 20 | 30 |
| | WQW | Si | 14 | 0.67 | 0 |
| | | $Si_{1-x}Ge_x$ | | 0.33 | 30 |
| | | Si | 1 | 0.67 | 0 |
| | Top Barrier | $Si_{1-x}Ge_x$ | 1 | 12 | 30 |
| | | | | | |
| C | Bottom Barrier | $Si_{1-x}Ge_x$ | 1 | 20 | 30 |
| | WQW | Si | 19 | 0.59 | 0 |
| | | $Si_{1-x}Ge_x$ | | 0.2 | 30 |
| | | Si | 1 | 0.59 | 0 |
| | Top Barrier | $Si_{1-x}Ge_x$ | 1 | 3 | 30 |
| | | | | | |

*kevin-peter.gradwohl@ikz-berlin.de

| Sample ID | Layer | Epilayer | Periods | Thickness (nm) | Ge-concentration (atm-%) |
|---|---|---|---|---|---|
| D | Bottom Barrier | $Si_{1-x}Ge_x$ | 1 | 20 | 30 |
| | Interlayer | Si | 1 | 5 | 0 |
| | WQW | Si | 48 | 0.49 | 0 |
| | | $Si_{1-x}Ge_x$ | | 0.33 | 18 |
| | | Si | 1 | 0.38 | 0 |
| | | | | | |
| E | Bottom Barrier | $Si_{1-x}Ge_x$ | 1 | 20 | 30 |
| | WQW | Si | 27 | 0.43 | 0 |
| | | $Si_{1-x}Ge_x$ | | 0.12 | 30 |
| | | Si | 1 | 0.43 | 0 |
| | Top Barrier | $Si_{1-x}Ge_x$ | 1 | 3 | 30 |

## SI2: XRR Limits in Lab-Based Measurements

Figure 1 summarizes several lab-based XRR measurements, performed with a Rigaku Smartlab system (cf. section *Online Methods* in main publication). In the corresponding graphs the WQW diffraction peaks (1st-, 2nd- and 3rd-order) are highlighted with arrows. The 1st-order diffraction peak of sample A and B can be clearly observed. For samples C, D and E it was not possible to observe any WQW diffraction peak in the lab-based measurements (curves not shown here).

In the context of the WQW investigation two other test samples were grown, here called ”test sample 1” and ”test sample 2”, with the same layer structure as sample A, B, C, and E. We would like to highlight here, that for test sample 1 it was possible to observe even a 2nd- and 3rd-order WQW diffraction peak in the lab-based XRR measurement.

The layer structure of test sample 2 is a duplicate of sample C (i.e., same growth settings, like source powers and shutter times). Compared to sample C, the WQW thickness was chosen to be approximately 80 nm, expecting to get more X-ray intensity from the WQW itself. In the corresponding XRR curve, the WQW diffraction peak can be observed. Longer measurement times (e.g., smaller step size and/or longer exposure times) did not improve the peak intensity significantly (orange curve), due to the correlated increase of background for longer exposure times. The measurement of sample C (no WQW peak detected) and test sample 2 (very weak WQW peak) gives a rough estimate for the lower detection limit of the WQW diffraction peak measured in lab-based XRR measurements.

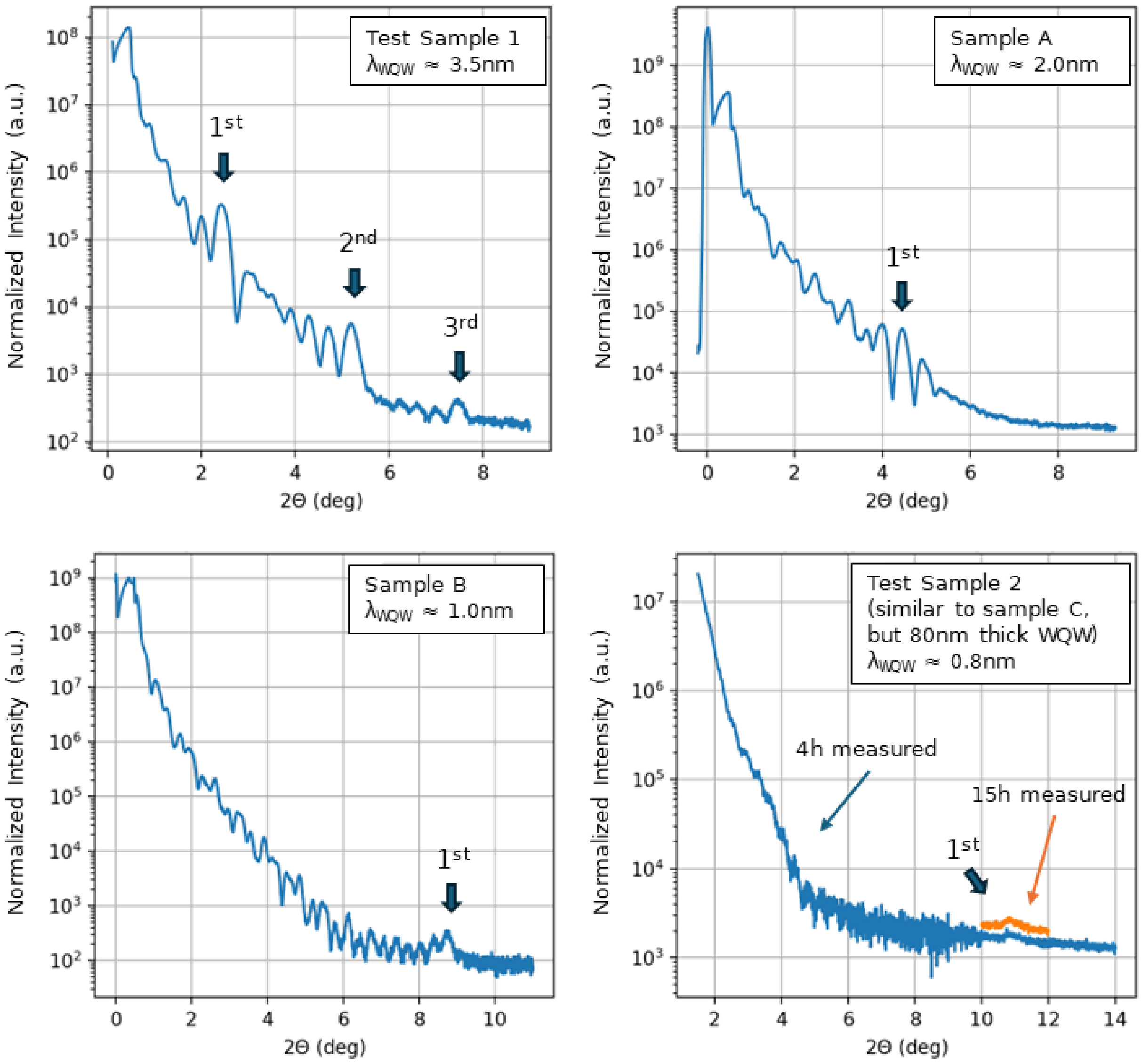


Figure 1: Different samples measured with lab-based X-ray setups (here: Rigaku Smartlab system), with Cu-K$\alpha$1 radiation ($\lambda_{\text{xray}}$ = 1.5406Å). Sample A and sample B are the samples discussed in the main publication. Sample 1 and sample 2 are test structures grown in context of the investigated samples of the main publication. The arrows indicate the WQW diffraction peaks. For sample 2 it becomes apparent, that the lab-based XRR measurement does not sufficiently resolve the WQW peak anymore, even when long measurement times are chosen.

## SI3: Limitations of X-Ray Reflectivity Measurements

The overall specular reflectivity, $I$ decreases with incidence angle $\theta$ as $I \propto sin(\theta)^{-4}$, while the WQW superlattice peak intensity increases to the density contrast $I \propto \Delta\rho$, and increases with the number of WQW periods $I \propto N^2$. Additionally $I$ decreases with increasing interface width $\sigma$ as $I \propto e^{-Q^2\sigma^2}$. Given that the scattering vector is $Q \propto sin(\theta)$, the intensity as a function of interface width is $I \propto e^{-sin(\theta)^2\sigma^2}$.

If one assumes a model WQW with a given constant Ge slope (limited by segregation), the realizable Ge modulation density amplitude is $\Delta\rho \propto \lambda_{\mathrm{WQW}}$, and the interface width between Ge modulations is $\sigma \propto \lambda_{\mathrm{WQW}}$ (assuming no other interface roughening, which likely is the case for very short periodicity WQWs). Furthermore, the number of WQW periods for a given QW thickness is $N \propto L^{-1}$.

If one put all together, and takes into account Bragg's law for diffraction on WQW density, $sin(\theta) \propto \lambda_{\mathrm{WQW}}^{-1}$, one can estimate the proportionality of the WQW diffraction peak intensity as a function of WQW period length $\lambda_{\mathrm{WQW}}$, as displayed in Tab. 2.

Table 2: Estimation of the WQW diffraction peak intensity with respect to the oscillation period $\lambda_{\mathrm{WQW}}$

| XRR intensity | WQW diffraction intensity |
|---|---|
| $I \propto sin(\theta)^{-4}$ | $I \propto \lambda_{\mathrm{WQW}}^{4}$ |
| $I \propto \Delta\rho$ | $I \propto \lambda_{\mathrm{WQW}}$ |
| $I \propto N^2$ | $I \propto \lambda_{\mathrm{WQW}}^{-2}$ |
| $I \propto e^{-sin(\theta)^2\sigma^2}$ | $I \propto e^{-\lambda_{\mathrm{WQW}}^{-2}\lambda_{\mathrm{WQW}}^{2}}$ |
| **total intensity:** | $I \propto \lambda_{\mathrm{WQW}}^{3}$ |

Consequently, the XRR WQW diffraction peak intensity is expected to be $I \propto \lambda_{\mathrm{WQW}}^{3}$. Given that 0.87 nm WQW were just detectable in synchroton X-ray based XRR measurements, limited by signal to noise ratio, one can estimate the necessary signal/noise reduction necessary to measure smaller period WQWs. To detect the smallest period WQWs presented here (0.49 nm), roughly half the periodicy of 0.87 nm, one would need around 8x improved signal-to-noise ratio. To strcuturally measure the so far unreachable 0.32 nm period WQWs (toughly a third of 0.87 nm), a signal/noise improvement of 27x would be required. While not completly in the realm of impossibility, a massive disproportional experimental effort (special experimental setup, helium-filled flight-tubes, ...) would be likely necessary. Furthermore, one should note that this is under the assumption of an ideal model WQW with constant Ge slopes where $\sigma \propto \lambda_{\mathrm{WQW}}$, but this will breaks down for small $\lambda_{\mathrm{WQW}}$ (e.g. due to the not perfectly atomically-flat nature of the growth surface morphology) and adds another exponential decay term to the intensity. In such extreme Wiggle Wells with low Ge concentration, the $E_{\mathrm{VS}}$ might be laterally mapped in fabricated quantum chips using conveyor-mode shuttling. As this is far beyond our scope here, we further focus here on GIXD investigation of WQW, a more suitable approach for thin layer stacks.

## SI4: XRR Simulations

Layer properties, such as the density $\rho$, thickness $t$ and interface roughness $\sigma$ can be derived from the measured XRR curves of sample A-C and sample E, by simulating the curves. In this regards, the simulation of the curves follows the recursive *Paratt* formalism [1]. Therefore, a model layerstack, consisting of $N$ individual layers $j$, is constructed in the first place. Each of the layers $j$ are attributed with the (input) layer parameters ($\rho_j$, $t_j$ and $\sigma_j$). In a second step, the Fresnel reflection coefficients $r_{j,j+1} = \frac{k_{z,j}-k_{z,j+1}}{k_{z,j}+k_{z,j+1}}$ of each layer $j$ are calculated, recursively (i.e., starting with the bottom most layer). $k_{z,j}$ are the vertical components of the (X-ray) wavevectors in the $j$-th layer, which are related to the incidence angle $\theta_i$ and the refractive index $n_j$ of the $j$-th layer via $k_{z,j} = \frac{2\pi}{\lambda_{\mathrm{xray}}}\sqrt{n_j^2 - \cos^2(\theta_i)}$. The Fresnel coefficient at the bottom most layer, i.e. the interface between substrate and $N$-th layer, is given by $R_j = \frac{r_{j,j+1}+R_{j+1}e^{2ik_{z,j+1}t_j}}{1+r_{j,j+1}R_{j+1}e^{2ik_{z,j+1}t_j}}$, due to the substrate reflectivity amplitude $R_N = 0$. The reflectivity intensity at the top interface is given by $R(q_z) = |R_0|^2$, eventually, where $q_z = \frac{4\pi}{\lambda}\sin(\theta_i)$. To account for the interface roughness $\sigma_j$ the Fresnel coefficients are modified with a Debye-Waller like factor, like $r_{z,j} \rightarrow r_{z,j}\exp -2k_{z,j}^2\sigma_j^2$.

Table 3: XRR simulation input and output layer structure/parameters of 2.0nm WQW (sample A)

| Layer | Simulation Input | | | Simulation Output | | |
|---|---|---|---|---|---|---|
| | Name | Composition | Individual/Coupled Layer Properties | Density (g/cm^3) | Thickness (A) | Roughness (A) |
| 0 | Air | N0.78O.21 | individual | 0.00125 | inf | 0 |
| 1 | Dirt | C | individual | 1.150713 | 9.727249 | 2.35757 |
| 2 | Oxide | SiO2 | individual | 4.917556 | 4.22427 | 8.760189 |
| 3 | Top Barrier | Si0.70Ge0.30 | individual | 3.326238 | 118.644349 | 7.328619 |
| 4 | QW_Si_start | Si | individual | 2.50952 | 14.303624 | 4.478816 |
| 5 | QW_SiGe1 | Si0.70Ge0.30 | | 3.113999 | 5.36006 | 3.334081 |
| 6 | QW_Si1 | Si | | 2.476994 | 13.981761 | 4.65477 |
| 7 | QW_SiGe2 | Si0.70Ge0.30 | coupled | 3.113999 | 5.369772 | 3.334081 |
| 8 | QW_Si2 | Si | (density, roughness) | 2.476994 | 14.093053 | 4.65477 |
| 9 | QW_SiGe3 | Si0.70Ge0.30 | | 3.113999 | 5.640497 | 3.334081 |
| 10 | QW_Si3 | Si | | 2.476994 | 14.083577 | 4.65477 |
| 11 | QW_SiGe_end | Si0.70Ge0.30 | individual | 3.177818 | 4.79218 | 3.238627 |
| 12 | QW_Si_end | Si | individual | 2.499843 | 15.883113 | 4.777651 |
| 13 | Bottom Barrier | Si0.70Ge0.30 | individual | 3.313259 | 211.703248 | 4.044641 |
| 14 | CMP | Si0.70Ge0.30 | individual | 3.115841 | 4.479729 | 4.251043 |
| 15 | Substrate | Si0.70Ge0.30 | individual | 3.329 | inf | 5.099904 |

With the recursive *Paratt* formalism the reflection intensity at the top interface for each incidence angle $\theta_i$ can be calculated, i.e., the simulation curve. Due to experimental discrepancies between the target layer parameters (density, thickness and interface roughness of each layer $j$), the simulation curve can differ from the measured curve. To improve the simulation curve to match the measured curve better, the recursive formalism can be implemented in a fitting procedure. Therefore, a range for each input parameter is defined (i.e., physical meaningful limits according to the epitaxial growth) and the simulation curve is fitted to the measured curve accordingly, by varying the input parameters within the individual ranges. In this regard, the input parameters ($\rho_j$, $t_j$, $\sigma_j$) can be either treated individually, or they can be coupled.

Samples A-C were simulated using a model layerstack with the individual WQW layer. Each layer within the layerstack has three input parameters: density, thickness and roughness. Thus, the complexity of the simulation scales with $3 \cdot N$, where $N$ is the number of individual layers in the stack. Whereas sample A consist of 4 oscillation periods (i.e. $N = 8$, within the WQW), sample B already consists of 14 (i.e. $N = 28$, within the WQW) periods and sample C even consists of 19 periods (i.e. $N = 38$, within the WQW). To reduce the simulation complexity of the WQWs, the density and roughness parameter were coupled for the individual WQW layers. Therefore, the simulation complexity was reduced to be $N + 2$, with only the thickness of each WQW layer being an individual parameter. Additionally, the interface layers between the $Si_{0.7}Ge_{0.3}$ barriers and the WQW were also treated with individual parameters, to account for any interfacial effects during the epitaxial growth. Those layers were called ”WQW-Si-start”, ”WQW-Si-end” and ”WQW-SiGe-end”.
Since, no WQW diffraction peak could be observed in the XRR curve of sample E, the sample was simulated with a ”simple” QW approach. Thus, the WQW was treated as a quantum well, i.e., a single layer, with an average Ge concentration.

In the Figures 2-5, the measurement data (black curves) and the simulation results (red curves) of sample A-C and sample E are shown. The corresponding simulation output parameters are given in Table 3-6.

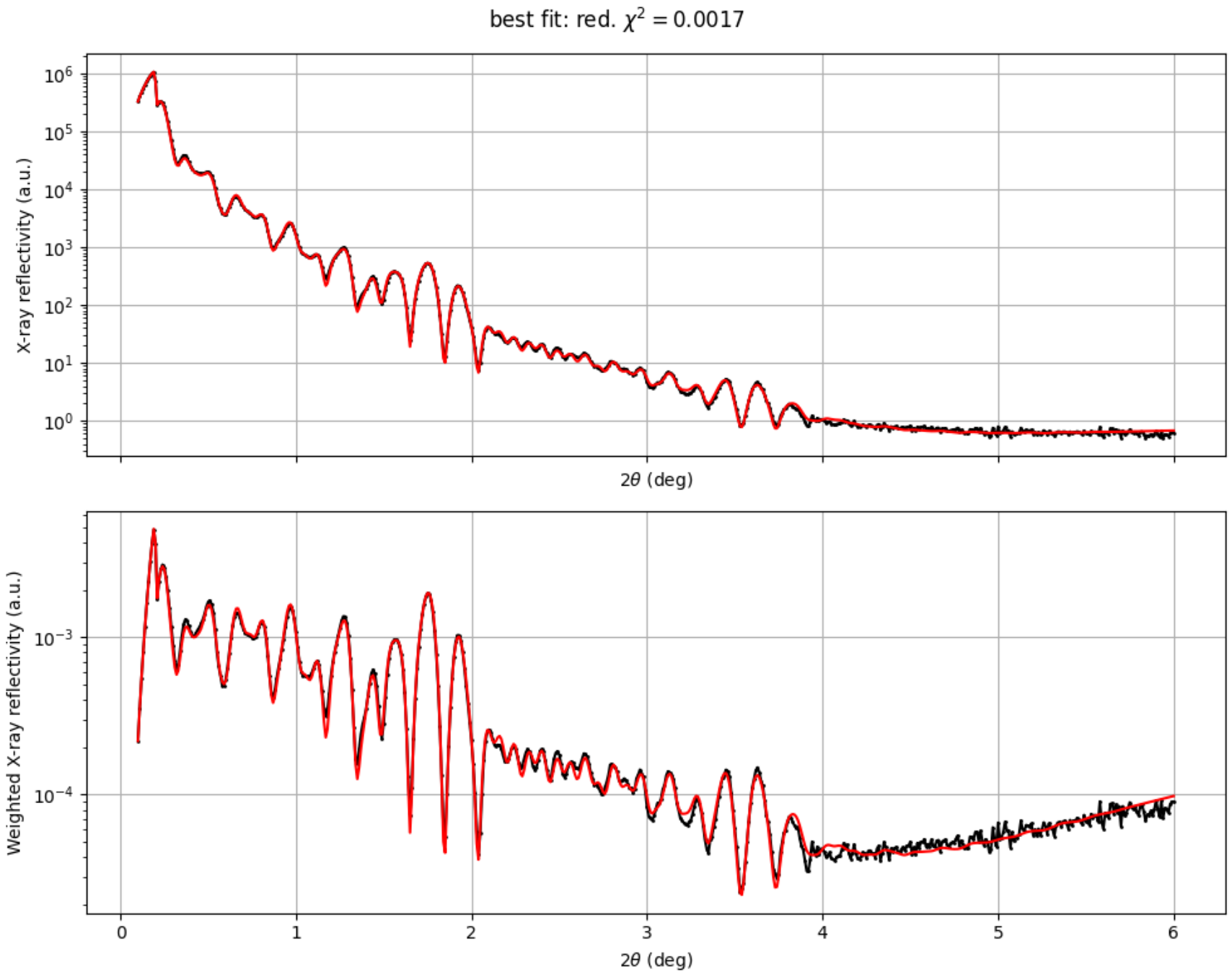


Figure 2: XRR Simulation results of 2.0nm WQW (sample A). The black curve is the measured data. The red curve the corresponding simulation curve, with the output layer parameters given in Tab. 3.

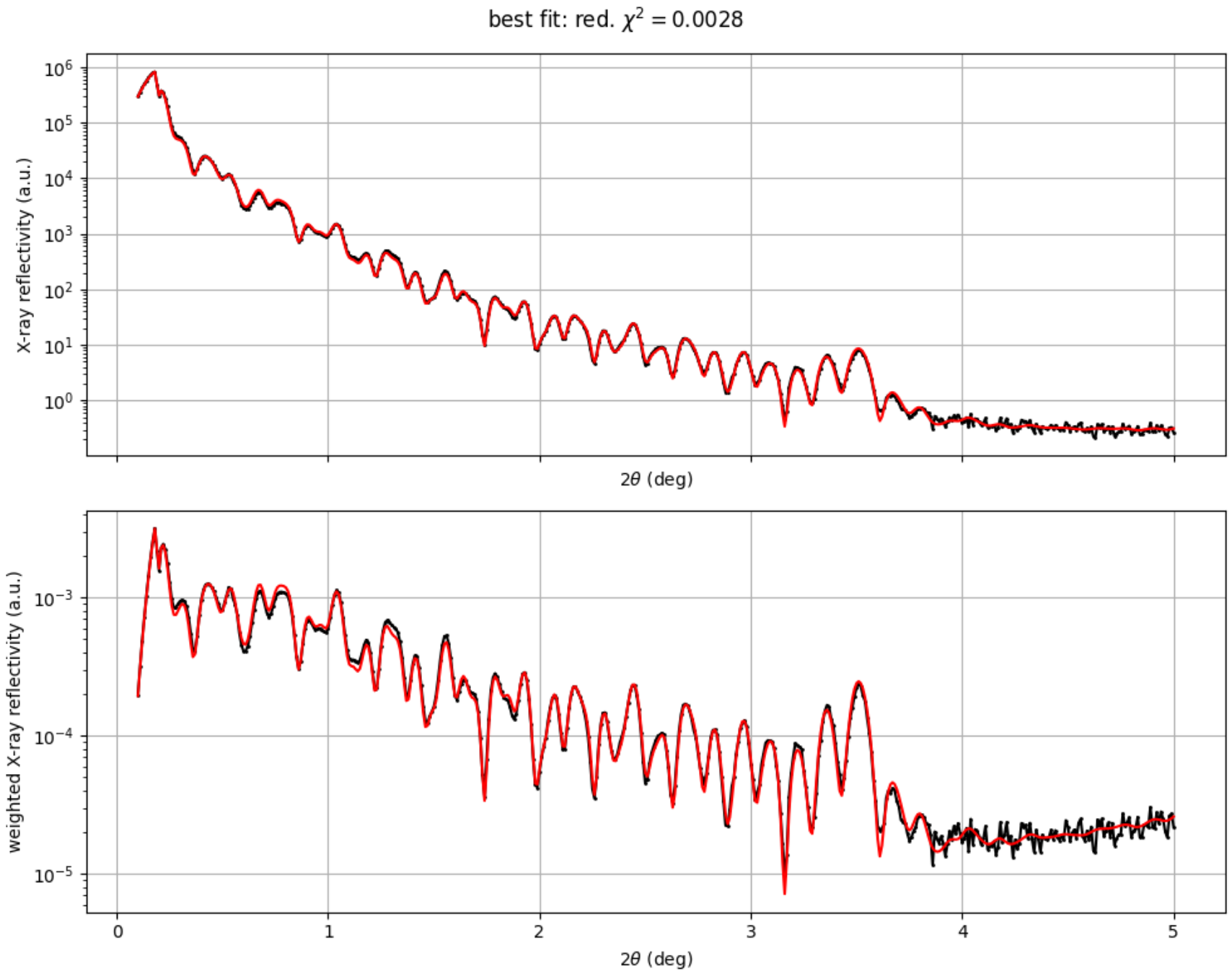


Figure 3: XRR Simulation results of 1.0nm (sample B). The black curve is the measured data. The red curve the corresponding simulation curve, with the output layer parameters given in Tab. 5.

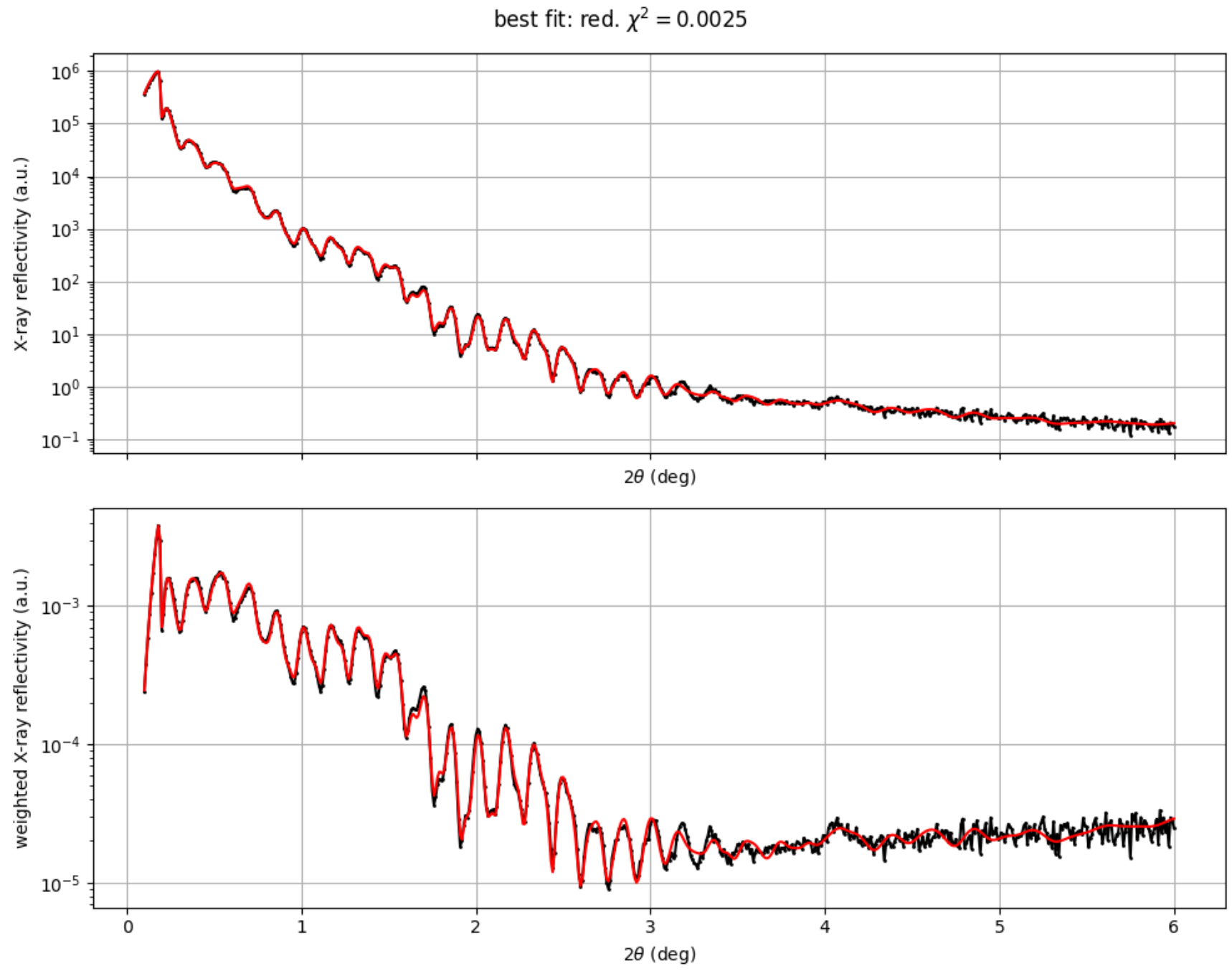


Figure 4: XRR Simulation results of 0.88nm WQW (sample C). The red curve the corresponding simulation curve, with the output layer parameters given in Tab. 6.

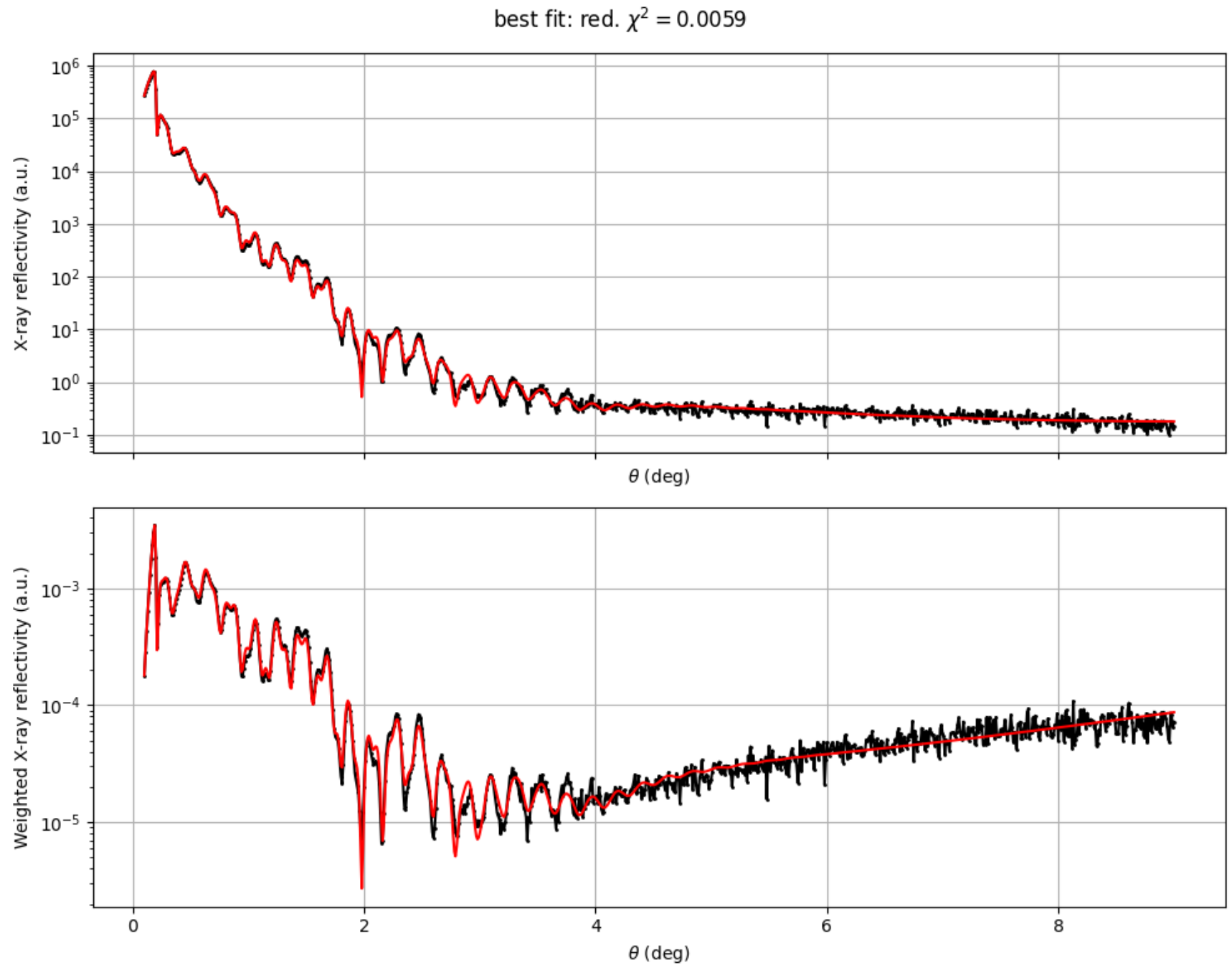


Figure 5: XRR Simulation results of 0.49nm (sample E). The red curve the corresponding simulation curve, with the output layer parameters given in Tab. 4.

Table 4: XRR simulation input and output layer structure/parameters of 0.49nm WQW (sample E)

| Layer | Simulation Input | | | Simulation Output | | |
|---|---|---|---|---|---|---|
| | Name | Composition | Individual/Coupled Layer Properties | Density (g/cm^3) | Thickness (A) | Roughness (A) |
| 1 | Dirt | C | individual | 1.954091 | 7.073361 | 3.542311 |
| 2 | Oxide | SiO2 | individual | 3.629166 | 0.095878 | 3.228982 |
| 3 | Top Barrier | Si0.70Ge0.30 | individual | 3.441401 | 30.165872 | 0.050027 |
| 4 | QW | Si0.85Ge0.15 | individual | 3.079101 | 135.072049 | 3.593002 |
| 5 | Bottom Barrier | Si0.70Ge0.30 | individual | 3.535349 | 228.197077 | 3.377266 |
| 6 | CMP | Si0.70Ge0.30 | individual | 2.191954 | 1.613369 | 6.715435 |
| 7 | Substrate | Si0.70Ge0.30 | individual | 3.432915 | inf | 5.581209 |

Table 5: XRR simulation input and output layer structure/parameters of 1.0nm WQW (sample B)

| Layer | Simulation Input | | | Simulation Output | | |
|---|---|---|---|---|---|---|
| | Name | Composition | Individual/Coupled Layer Properties | Density (g/cm^3) | Thickness (A) | Roughness (A) |
| 1 | Dirt | C | individual | 0.1 | 14.999997 | 2.892955 |
| 2 | Oxide | SiO2 | individual | 4.527118 | 2.977752 | 5.152899 |
| 3 | Top Barrier | Si0.70Ge0.30 | individual | 3.325226 | 118.419799 | 2.762014 |
| 4 | QW_Si_start | Si | individual | 2.51553 | 6.821056 | 4.248641 |
| 5 | QW_SiGe1 | Si0.70Ge0.30 | coupled (density, roughness) | 2.823093 | 8 | 0.964541 |
| 6 | QW_Si1 | Si | | 2.540677 | 2 | 0.983418 |
| 7 | QW_SiGe2 | Si0.70Ge0.30 | | 2.823093 | 6.982143 | 0.964541 |
| 8 | QW_Si2 | Si | | 2.540677 | 2.70502 | 0.983418 |
| 9 | QW_SiGe3 | Si0.70Ge0.30 | | 2.823093 | 7.809507 | 0.964541 |
| 10 | QW_Si3 | Si | | 2.540677 | 2.880186 | 0.983418 |
| 11 | QW_SiGe4 | Si0.70Ge0.30 | | 2.823093 | 4.999719 | 0.964541 |
| 12 | QW_Si4 | Si | | 2.540677 | 6.712607 | 0.983418 |
| 13 | QW_SiGe5 | Si0.70Ge0.30 | | 2.823093 | 4.557299 | 0.964541 |
| 14 | QW_Si5 | Si | | 2.540677 | 3.724305 | 0.983418 |
| 15 | QW_SiGe6 | Si0.70Ge0.30 | | 2.823093 | 7.253629 | 0.964541 |
| 16 | QW_Si6 | Si | | 2.540677 | 3.289954 | 0.983418 |
| 17 | QW_SiGe7 | Si0.70Ge0.30 | | 2.823093 | 5.889222 | 0.964541 |
| 18 | QW_Si7 | Si | | 2.540677 | 4.277566 | 0.983418 |
| 19 | QW_SiGe8 | Si0.70Ge0.30 | | 2.823093 | 6.552628 | 0.964541 |
| 20 | QW_Si8 | Si | | 2.540677 | 3.896257 | 0.983418 |
| 21 | QW_SiGe9 | Si0.70Ge0.30 | | 2.823093 | 5.220218 | 0.964541 |
| 22 | QW_Si9 | Si | | 2.540677 | 4.99086 | 0.983418 |
| 23 | QW_SiGe10 | Si0.70Ge0.30 | | 2.823093 | 5.388457 | 0.964541 |
| 24 | QW_Si10 | Si | | 2.540677 | 5.084331 | 0.983418 |
| 25 | QW_SiGe11 | Si0.70Ge0.30 | | 2.823093 | 4.263607 | 0.964541 |
| 26 | QW_Si11 | Si | | 2.540677 | 6.011198 | 0.983418 |
| 27 | QW_SiGe12 | Si0.70Ge0.30 | | 2.823093 | 4.548052 | 0.964541 |
| 28 | QW_Si12 | Si | | 2.540677 | 4.510666 | 0.983418 |
| 29 | QW_SiGe13 | Si0.70Ge0.30 | | 2.823093 | 7.517519 | 0.964541 |
| 30 | QW_Si13 | Si | | 2.540677 | 2.723035 | 0.983418 |
| 31 | QW_SiGe_end | Si0.70Ge0.30 | individual | 2.880483 | 8 | 0.758526 |
| 32 | QW_Si_end | Si | individual | 2.446928 | 3.194021 | 2.19419 |
| 33 | Bottom Barrier | Si0.70Ge0.30 | individual | 3.344477 | 200.370458 | 3.492336 |
| 34 | CMP | Si0.70Ge0.30 | individual | 2.988541 | 3.189162 | 6.768374 |
| 35 | Substrate | Si0.70Ge0.30 | individual | 3.329 | inf | 4.129865 |

Table 6: XRR simulation input and output layer structure/parameters of 0.88nm WQW (sample C)

| Layer | Simulation Input | | | Simulation Output | | |
|---|---|---|---|---|---|---|
| | Name | Composition | Individual/Coupled Layer Properties | Density (g/cm^3) | Thickness (A) | Roughness (A) |
| 1 | Dirt | C | individual | 0.174605 | 14.999999 | 2.57996 |
| 2 | Oxide | SiO2 | individual | 3.043983 | 8.481201 | 5.911314 |
| 3 | Top Barrier | Si0.70Ge0.30 | individual | 3.256034 | 27.34745 | 2.081818 |
| 4 | QW_Si_start | Si | individual | 2.632759 | 2.00977 | 5.5 |
| 5 | QW_SiGe1 | Si0.70Ge0.30 | coupled (density, roughness) | 2.755414 | 4.259191 | 2.091827 |
| 6 | QW_Si1 | Si | | 2.470314 | 2.000173 | 0.98293 |
| 7 | QW_SiGe2 | Si0.70Ge0.30 | | 2.755414 | 0.5 | 2.091827 |
| 8 | QW_Si2 | Si | | 2.470314 | 6.785876 | 0.98293 |
| 9 | QW_SiGe3 | Si0.70Ge0.30 | | 2.755414 | 1.852659 | 2.091827 |
| 10 | QW_Si3 | Si | | 2.470314 | 6.060901 | 0.98293 |
| 11 | QW_SiGe4 | Si0.70Ge0.30 | | 2.755414 | 2.203709 | 2.091827 |
| 12 | QW_Si4 | Si | | 2.470314 | 7.503993 | 0.98293 |
| 13 | QW_SiGe5 | Si0.70Ge0.30 | | 2.755414 | 1.4594 | 2.091827 |
| 14 | QW_Si5 | Si | | 2.470314 | 7.467186 | 0.98293 |
| 15 | QW_SiGe6 | Si0.70Ge0.30 | | 2.755414 | 3.709768 | 2.091827 |
| 16 | QW_Si6 | Si | | 2.470314 | 6.669933 | 0.98293 |
| 17 | QW_SiGe7 | Si0.70Ge0.30 | | 2.755414 | 1.871127 | 2.091827 |
| 18 | QW_Si7 | Si | | 2.470314 | 10 | 0.98293 |
| 19 | QW_SiGe8 | Si0.70Ge0.30 | | 2.755414 | 1.438295 | 2.091827 |
| 20 | QW_Si8 | Si | | 2.470314 | 5.414189 | 0.98293 |
| 21 | QW_SiGe9 | Si0.70Ge0.30 | | 2.755414 | 2.604721 | 2.091827 |
| 22 | QW_Si9 | Si | | 2.470314 | 5.294367 | 0.98293 |
| 23 | QW_SiGe10 | Si0.70Ge0.30 | | 2.755414 | 1.462411 | 2.091827 |
| 24 | QW_Si10 | Si | | 2.470314 | 4.483226 | 0.98293 |
| 25 | QW_SiGe11 | Si0.70Ge0.30 | | 2.755414 | 2.951669 | 2.091827 |
| 26 | QW_Si11 | Si | | 2.470314 | 6.694964 | 0.98293 |
| 27 | QW_SiGe12 | Si0.70Ge0.30 | | 2.755414 | 2.280551 | 2.091827 |
| 28 | QW_Si12 | Si | | 2.470314 | 6.195778 | 0.98293 |
| 29 | QW_SiGe13 | Si0.70Ge0.30 | | 2.755414 | 0.867236 | 2.091827 |
| 30 | QW_Si13 | Si | | 2.470314 | 7.24156 | 0.98293 |
| 31 | QW_SiGe14 | Si0.70Ge0.30 | | 2.755414 | 0.892241 | 2.091827 |
| 32 | QW_Si14 | Si | | 2.470314 | 8.250455 | 0.98293 |
| 33 | QW_SiGe15 | Si0.70Ge0.30 | | 2.755414 | 0.790236 | 2.091827 |
| 34 | QW_Si15 | Si | | 2.470314 | 7.326679 | 0.98293 |
| 35 | QW_SiGe16 | Si0.70Ge0.30 | | 2.755414 | 2.770139 | 2.091827 |
| 36 | QW_Si16 | Si | | 2.470314 | 8.73495 | 0.98293 |
| 37 | QW_SiGe17 | Si0.70Ge0.30 | | 2.755414 | 2.524604 | 2.091827 |
| 38 | QW_Si17 | Si | | 2.470314 | 8.808793 | 0.98293 |
| 39 | QW_SiGe18 | Si0.70Ge0.30 | | 2.755414 | 4.596162 | 2.091827 |
| 40 | QW_Si18 | Si | | 2.470314 | 7.484153 | 0.98293 |
| 41 | QW_SiGe_end | Si0.70Ge0.30 | individual | 2.892828 | 2.232855 | 4.106006 |
| 42 | QW_Si_end | Si | individual | 2.616471 | 9.803822 | 0.100002 |
| 43 | Bottom Barrier | Si0.70Ge0.30 | individual | 3.300321 | 244.951681 | 3.169345 |
| 44 | CMP | Si0.70Ge0.30 | individual | 2.902325 | 1.381158 | 6.019372 |
| 45 | Substrate | Si0.70Ge0.30 | individual | 3.329 | inf | 4.818413 |

## SI5: XRR Ge Concentration Profile Analysis

The XRR simulation results, i.e., the density, roughness and thickness of each layer, given in the previous section allow for a more detailed analysis of the corresponding WQW. Due to the fact that only Si and Ge were present during the epitaxial growth of the layers, the density of the epitaxial layers can be directly converted into a Ge concentration, using Vegard's Law

$$\rho(x) = 2.329 + 3.493x - 0.499x^2, \tag{1}$$

with $x$ being the atomic fraction of Ge in $Si_{1-x}Ge_x$. Taking the (interface) roughness and thickness parameters into account, Ge concentration profiles can be deduced from the simulation parameters. In Figures 6-8 the Ge profiles within the QW are illustrated for sample A-C. The Ge profiles were slightly adjusted on the y-axis to match a targeted (bottom) barrier concentration of $x = 0.3$. This minor change of the absolute Ge concentration does not change the discussed results in the main manuscript (except a minor change in the random alloy-disorder contribution to $E_{\mathrm{VS}}$) and was solely applied for better comparison between the different WQW samples.
For further investigation, the peak maxima and minima were extracted from the profiles, and highlighted as red and blue points. In case of non-ideal sinusoidal (e.g., plateaus) oscillations, the plateau edges were extracted, and highlighted by the red and blue vertical lines. The cyan and yellow lines show the calculated (linear) slopes. The calculated slopes for the Si-SiGe and SiGe-Si interfaces are summarized in Table 2 in the main manuscript, where the slope is given by the mean value and the uncertainty given by its standard deviation. To account for interfacial effects, some of the WQW slopes are neglected (cf. SI4).

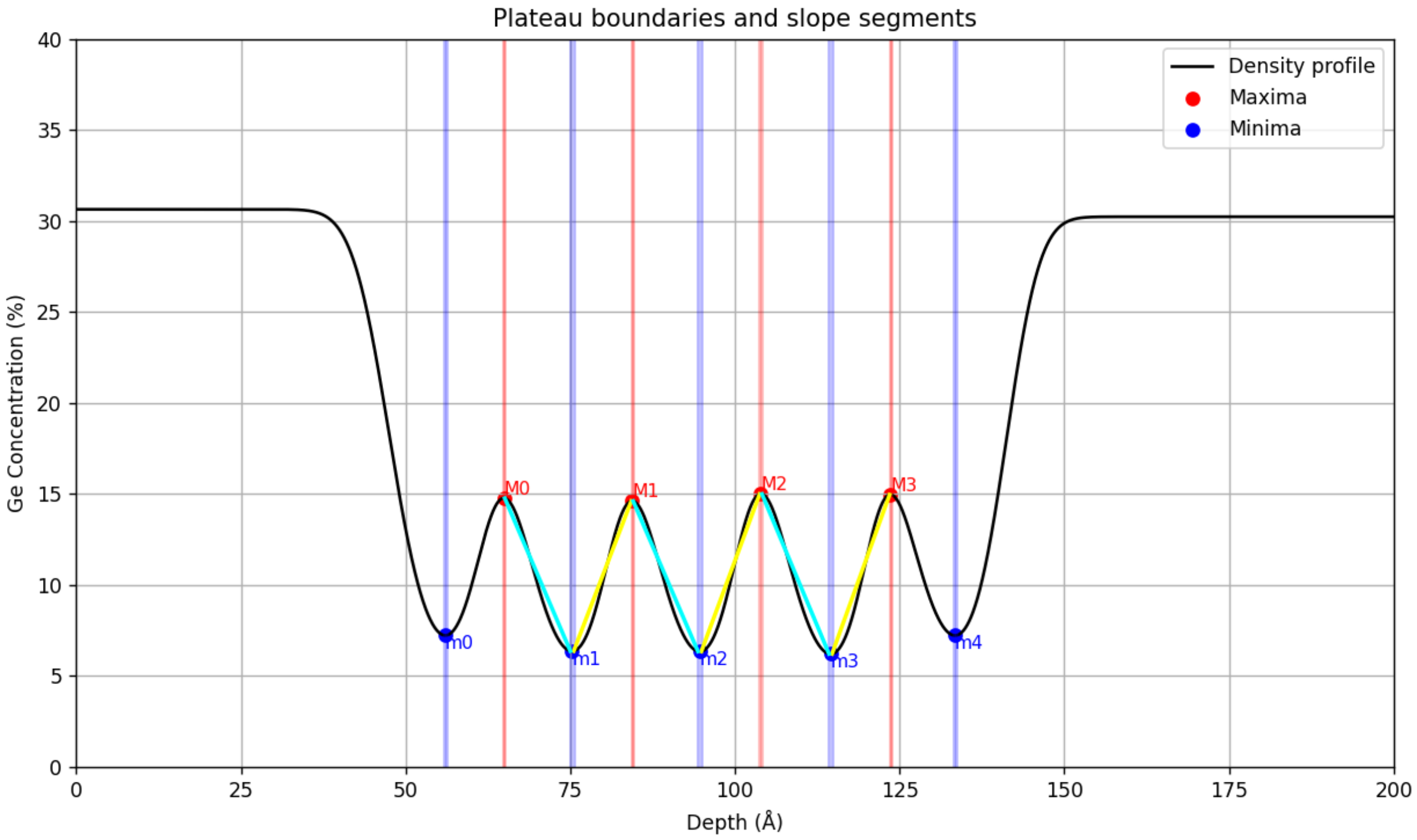


Figure 6: Ge concentration profile derived from XRR simulation results of 2.0nm sample A

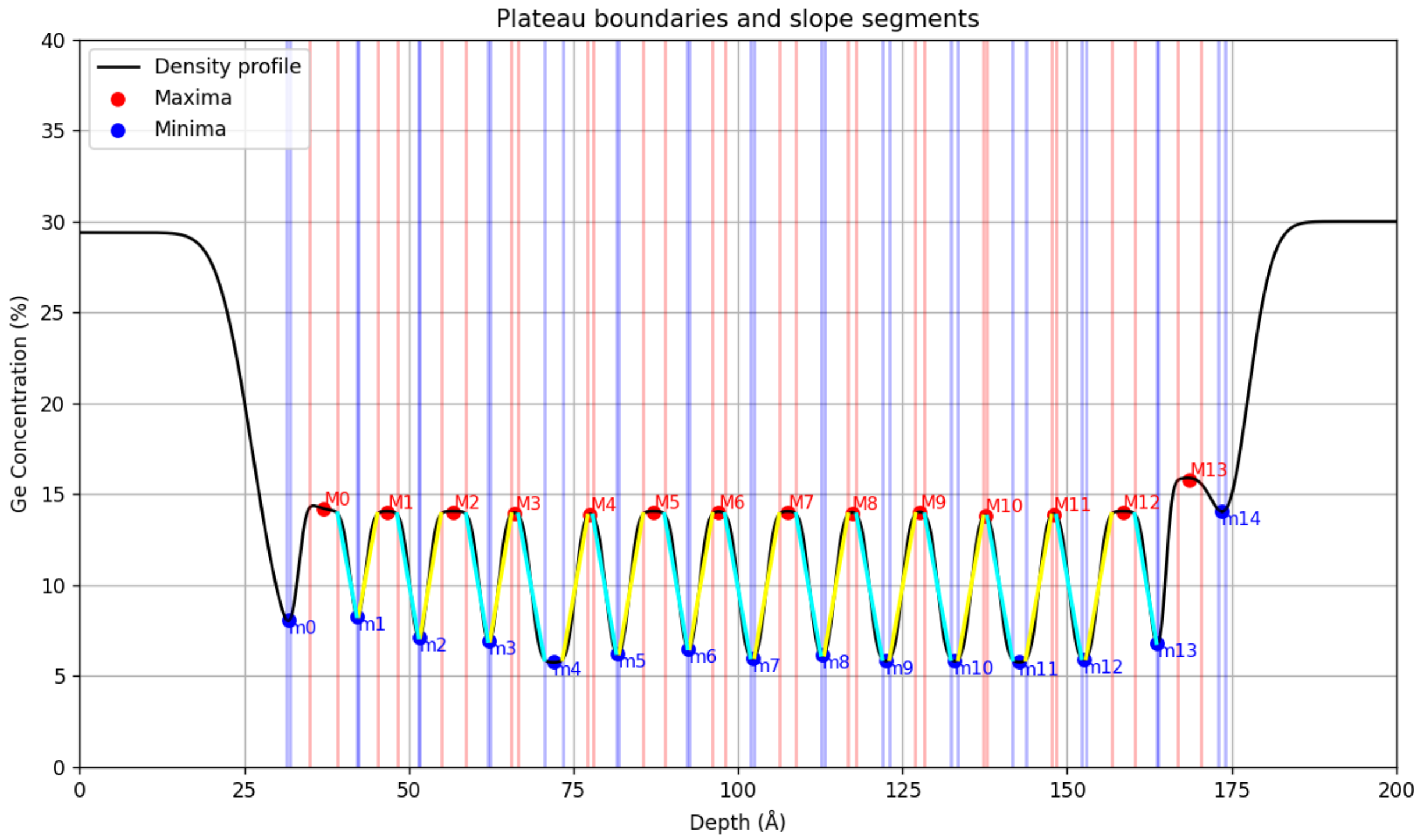


Figure 7: Ge concentration profile derived from XRR simulation results of 1.02nm sample B

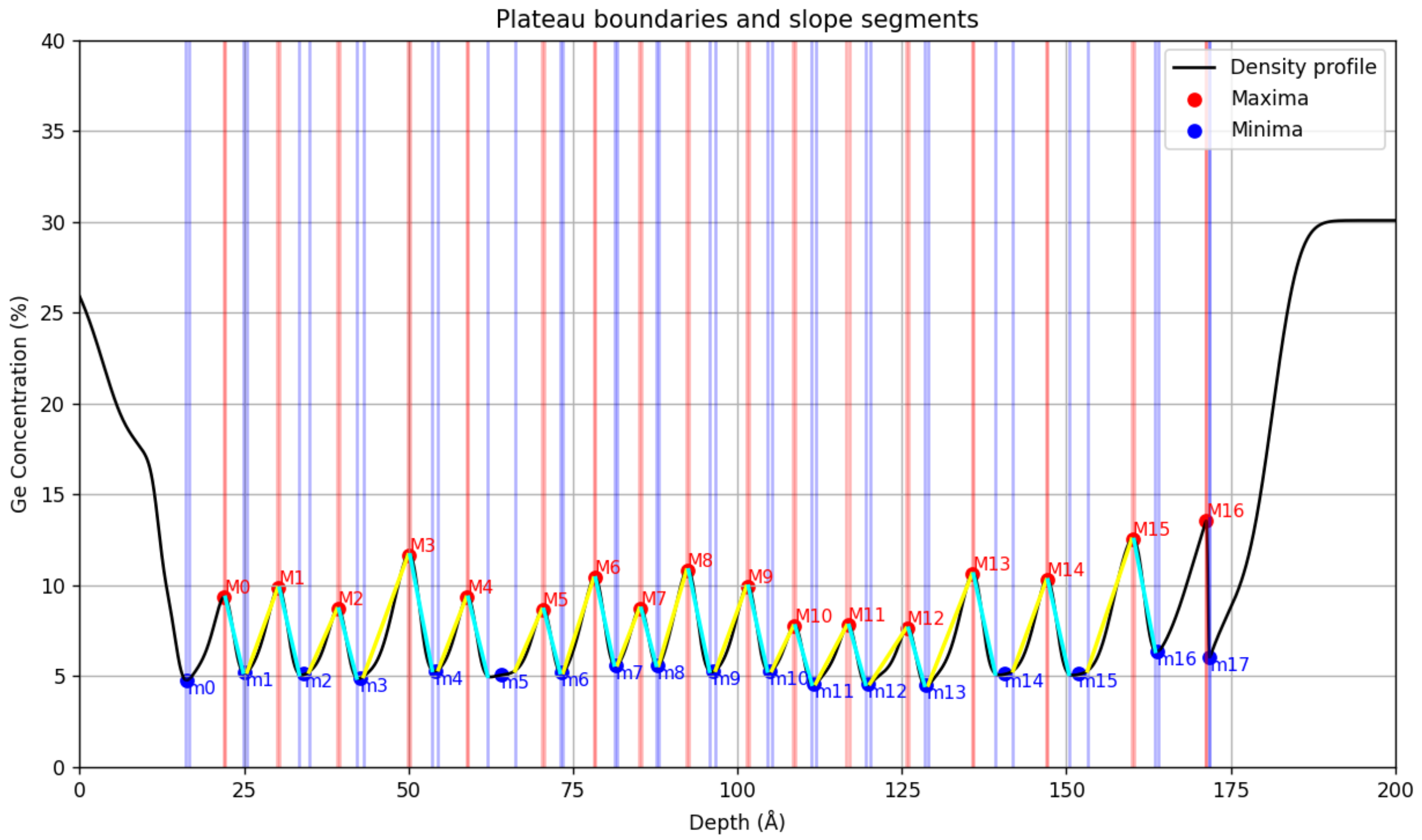


Figure 8: Ge concentration profile derived from XRR simulation results of 0.88nm sample C

## SI6: STEM Data Analysis

Fig. 9 shows the derived Ge concentration from the STEM-EDS measurement (green curve). For comparison, the derived profile from the XRR simulation is shown additionally (blue curve). A difference of the absolute Ge concentration of both curves is observed. This can be traced back to calibration issues of the EDS measurement and to the differences of the underlying physical principles of both measurements (e.g., interaction of X-rays/electrons with the material and lateral dimension of measured area). For better comparison of both measurements, the STEM-EDS curve was calibrated (cyan curve) to match the Ge concentration of the bottom barrier in the XRR simulation curve. For this purpose, the curve was scaled linearly, with $\rho_{\text{scaled}} = (\rho_{\text{EDS}}) \cdot 1.65 - 7$, as a first approximation. From the calibrated curve, the maxima and minima were detected and the slopes extracted (cf. Fig. 10).

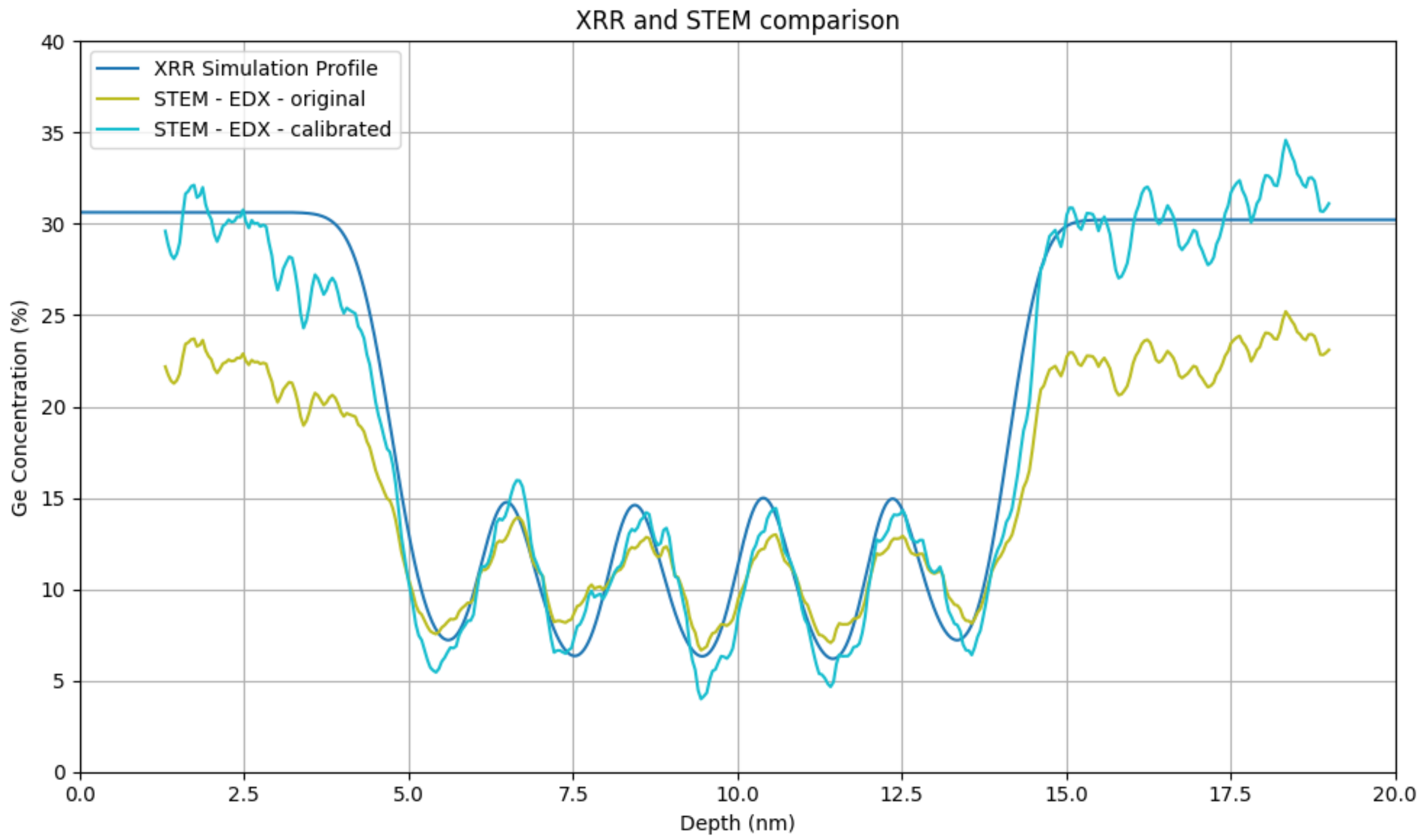


Figure 9: Ge concentration of Sample A derived from XRR simulation. Additionally, the extracted Ge concentration from the STEM-EDX measurement is illustrated. In a first step (cyan curve) the EDX curve was (slightly) adjusted/calibrated to match the $Si_{0.7}Ge_{0.3}$ ratio in the bottom barrier.

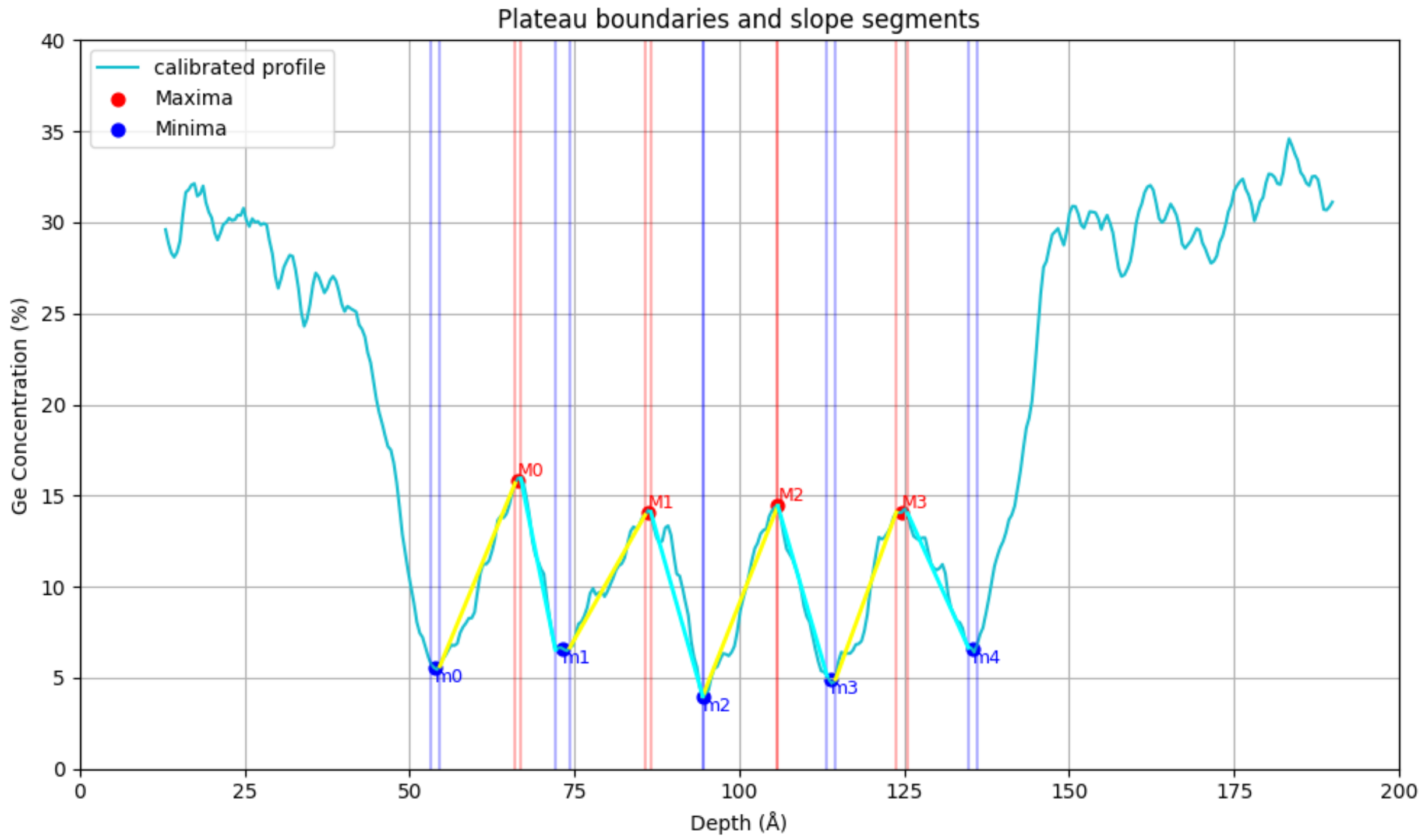


Figure 10: Ge concentration profile of sample A (2.0 nm-WQW) derived from the STEM-EDS measurement. The maxima and minima are highlighted in blue and red. Due to the deviation from an ideal sinusoidal shape of the oscillation curve, plateau boundaries were included for a more precise slope calculation. The slopes used for the mean (and standard deviation) are shown as cyan and yellow lines.

## SI7: Estimation of Ge Concentration Profile for Sample D and E

As described in the main manuscript, it was not possible to detect a WQW peak in the XRR measurements of sample D and E, due to the limited signal-to-noise ratio. Therefore, it is not possible to extract physical meaningful WQW layer parameters (i.e., density, roughness and thickness) from the simulations of the corresponding XRR measurements of sample D and E.
Nevertheless, the WQW periodicity $\lambda_{\mathrm{WQW}}$ could be extracted from the GIXD CTR measurements. WQW periodicities of (0.62±0.01) nm and (0.49±0.02) nm for sample D and E, respectively, were extracted. The individual layer thickness of the Si $t_{\mathrm{WQW,Si}}$ and SiGe $t_{\mathrm{WQW,SiGe}}$ layers can not be determined exactly. For simplicity, we assume here an equal layer thickness $t_{\mathrm{WQW,Si}} = t_{\mathrm{WQW,SiGe}} = t = \frac{1}{2}\lambda_{\mathrm{WQW}}$. Another simplification is done by choosing the (interface) roughness of the layers to be $\rho = \frac{1}{2}t$. This approximation is considered to be reasonable, comparing the (interface) roughnesses of the XRR simulation of sample B and C, which are in the range of $\rho \approx 1 - 2$ Å.

Considering both assumptions (layer thicknesses and roughness), the Ge amplitude of the WQWs can be estimated from the signal-to-noise ratio of the main diffraction peak relative to the WQW satellite peaks. For this purpose, the web simulation tool on *Sergey Stephanov's X-ray Server* (https://x-server.gmca.aps.anl.gov/) was used for simulating the CTR curves in a rudimentary way (i.e., only investigating main and WQW satellite peak intensity, neglecting discrepancies in the overall appearance of the simulation curve with respect to the measured curve). In the web simulation tool, the case for "coplanar extremely asymmetric diffraction of synchrotron radiation" (GID_SL) was chosen. By varying the Ge concentration amplitudes in the simulated layerstacks and comparing the the main peak to WQW peak intensity, Ge amplitudes of 8 at-% and 5 at-% for sample D and E, respectively, were extracted.

In Figures 11 and 12 the corresponding estimated Ge concentration profile is shown for sample D and E, respectively.

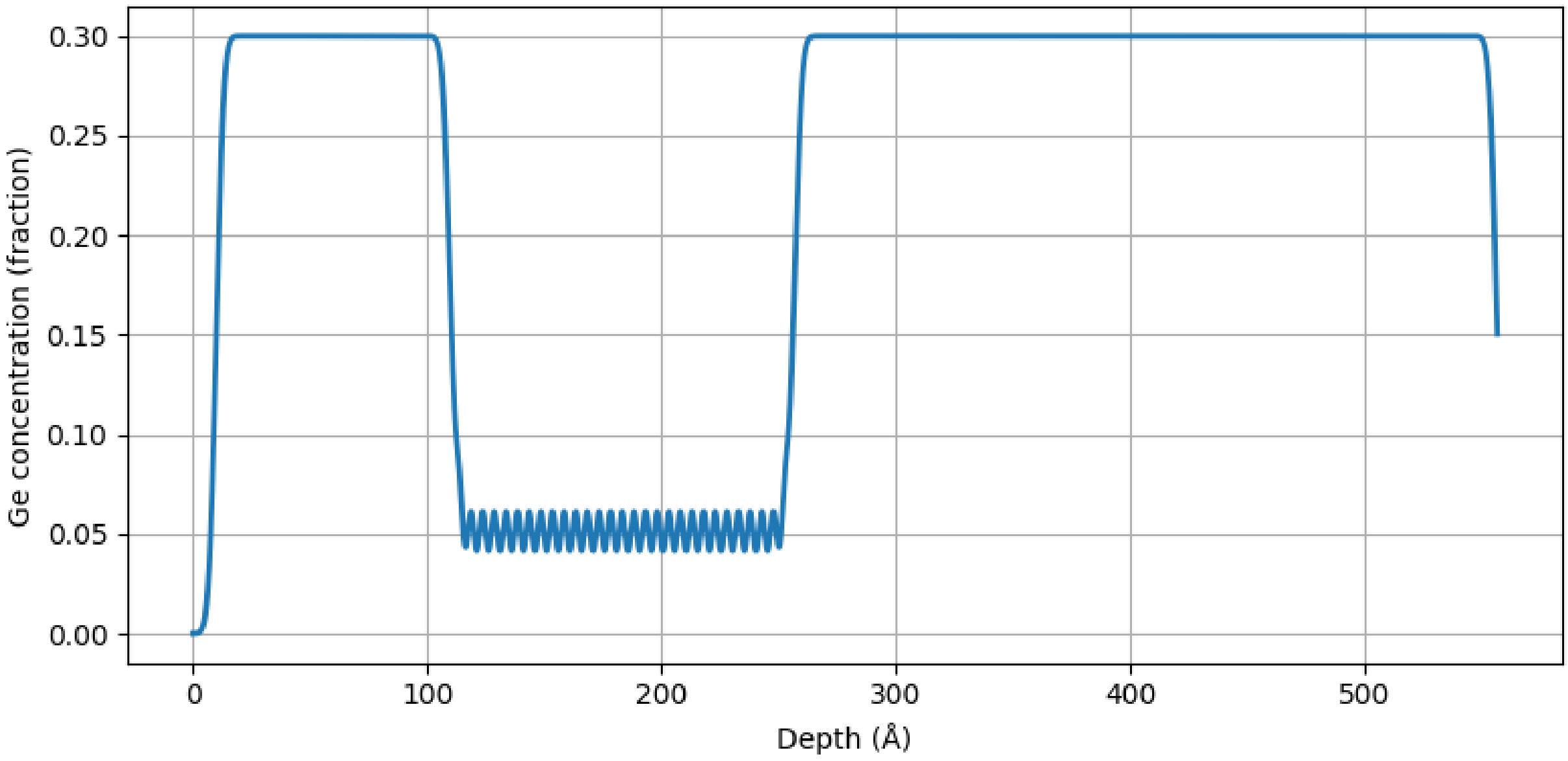


Figure 12: Estimated Ge concentration profile of sample E (0.49nm-WQW).

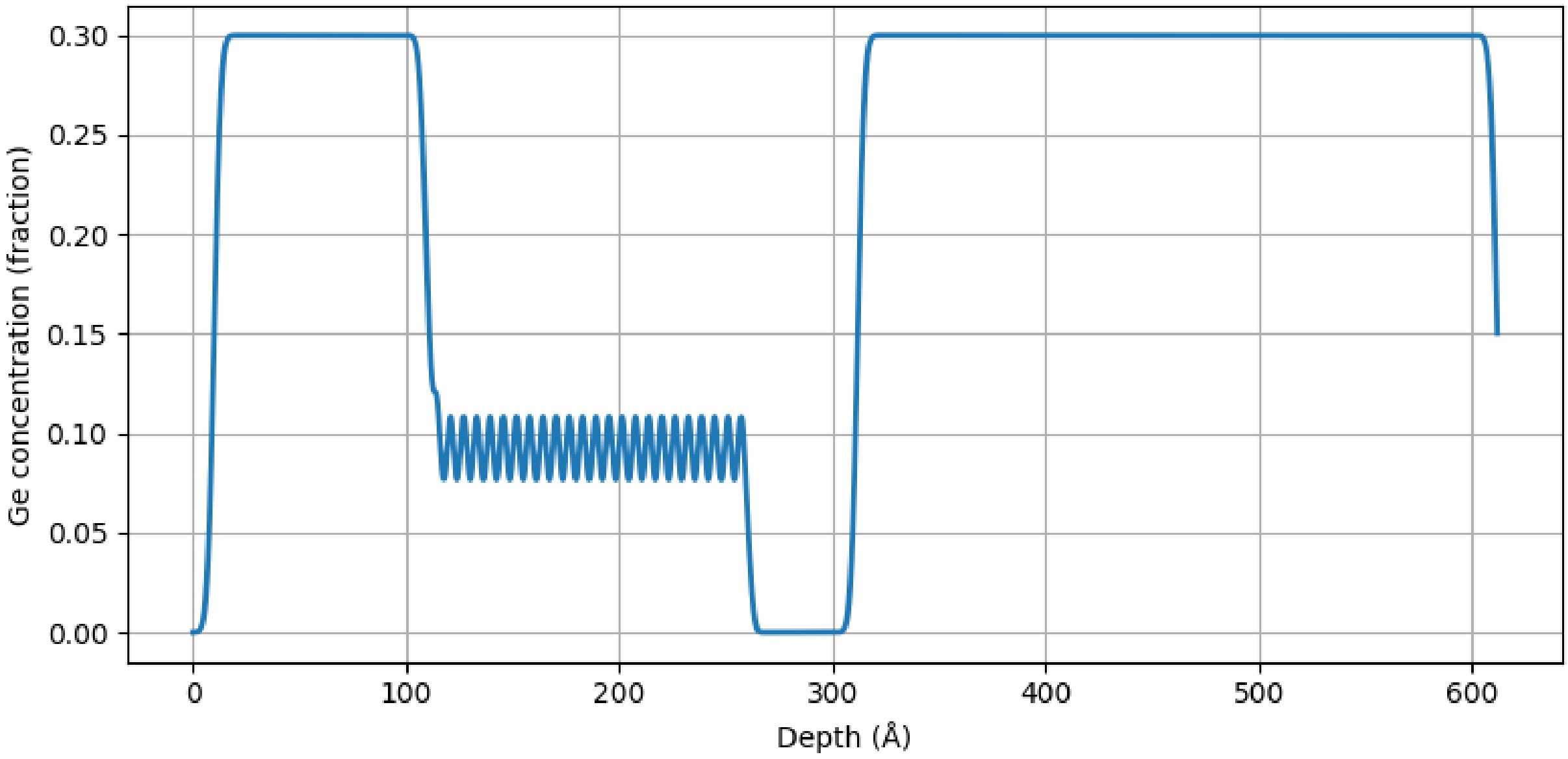


Figure 11: Estimated Ge concentration profile of sample D (0.62nm-WQW). Sample D is the only sample of the sample series with a slightly different epitaxial layer structure. Compared to sample A-C and E, it also includes a 5 nm Si layer between the $Si_{0.7}Ge_{0.3}$ bottom barrier and WQW (cf. target sample structures in Tab. 1). For the valley splitting energy simulation, a $Si_{0.7}Ge_{0.3}$ top barrier was included in the given profile.

## SI8: Valley Splitting Simulations

In this section, we provide additional details about the valley splitting in sample B and about the engineering of "trapezoidal" wiggle wells.

We first plot in Fig. 13 the valley splittings computed in sample B as a function of the dot radius[1] $r_{\|}$. The vertical electric field is $E_z = -5$ mV/nm, so that the electron is localized toward the bottom interface of the well (for reasons that will be discussed hereafter). The data for sample B are compared to the valley splittings calculated in the plain quantum well with same average Ge content. As discussed in the main text, the valley splittings in plain quantum wells are opened by alloy disorder, which averages out when the radius of the dot increases [2]. The median valley splitting $\overline{E_{\rm vs}}$ and inter-quartile range (IQR) of plain quantum wells are thus comparable and decrease approximately as $1/r_{\|}$. On the opposite, the median valley splitting of sample B is much larger than the IQR (which still decreases as $\approx 1/r_{\|}$) and is almost independent on dot size ($\overline{E_{\rm vs}} \approx 950\,\mu$eV) once $r_{\|} \gtrsim 10$ nm. This is the signature of a deterministic enhancement of the valley splittings by the wiggle well structure. The calculated median valley splitting appears to decrease again when $r_{\|} \gtrsim 18$ nm as the first excited state (used in the statistics) can be an orbital (dotted line) rather than a valley excitation.

We can rationalize the behavior of samples A-E using the "$2k_0$" theory [2, 3]. In this approximation, the valley splitting reads

$$E_{\rm vs} = 2\left|\int dz\, e^{-2ik_0 z}\, V(z)\, |\psi(z)|^2\right| , \tag{2}$$

where $\psi(z)$ is the electron envelope function and $V(z) \propto c_{\rm Ge}(z)$ is an inter-valley potential proportional to the Ge concentration $c_{\rm Ge}(z)$. In other words, $E_{\rm vs}$ is proportional to the Fourier transform $|\mathcal{F}(q)|$ of $f(z) = c_{\rm Ge}(z)|\psi(z)|^2$ at wave number $q = 2k_0$. This expression neglects alloy disorder and valley-orbit mixing, but is expected to capture the main trends in the deterministic regime where the valley splitting is dominated by the modulations of $c_{\rm Ge}(z)$.

We can complete the above integration for the Ge concentration profiles extracted from the experiments using a model wave function $\psi(z)$. For that purpose, we assume confinement in a $Si_{0.7}Ge_{0.3}$/Si/$Si_{0.7}Ge_{0.3}$ quantum well with thickness $L = 11$ nm at vertical electric field $E_z = -5$ mV/nm. The functions $f(z)$ and $|\mathcal{F}(q)|$ of samples A-E are plotted[2] in Fig. 14. As discussed in the main text, sample B shows the largest Fourier component at $q = 2k_0$, because its steeper, non-sinusoidal WQW modulations with wave length $\lambda_{\rm WQW} \approx 1.02$ nm give rise to strong harmonics (the envelope of the Fourier transform being expected to decrease as $1/q$ for perfectly square, periodic modulations). The third harmonic at $q \approx 2k_0$, broadened by the squared envelope function, is actually responsible for the enhancement of the valley splittings.

To get a better understanding of sample B, we can consider the limiting case of an ideal "trapezoidal $2k_0/3$ WQW" with perfectly periodic, trapezoidal modulations of the Ge concentration at wave length $\lambda_{\rm WQW} = 3\pi/k_0$. Such a test structure (Fig. 15a) alternates layers of Si and SiGe with the same thickness $t = \lambda_{\rm WQW}/2 \approx 5$ Å. The Ge concentration rises or falls linearly with finite slope $s$ at the interfaces between these layers. The amplitude $|C_n|$ of the harmonics of such a trapezoidal modulation is[3]

$$|C_n| = c_{\rm WQW}\left|{\rm sinc}\left(\frac{n\pi}{2}\right){\rm sinc}\left(\frac{n\pi c_{\rm WQW}}{s\lambda_{\rm WQW}}\right)\right| , \tag{3}$$

where ${\rm sinc}(x) = \sin(x)/x$. This expression nicely illustrates the dependence of $|C_n|$ on the slope $s$ of the edges; when $s \to \infty$ (square profile), $|C_n| \propto 1/n$, and when $s = 2c_{\rm WQW}/\lambda_{\rm WQW}$ (triangular profile), $|C_n| \propto 1/n^2$. In particular, the amplitude of the third harmonic at $q = 2k_0$ is

$$|C_3| = \frac{2c_{\rm WQW}}{3\pi}\left|{\rm sinc}\left(\frac{c_{\rm WQW}k_0}{s}\right)\right| = \frac{2s}{3\pi k_0}\left|\sin\left(\frac{c_{\rm WQW}k_0}{s}\right)\right| . \tag{4}$$

Strikingly, $|C_3|$ oscillates with the Ge concentration $c_{\rm WQW}$ in the wiggles (for a given slope $s$). It is maximum when $c_{\rm WQW} = (2m+1)\pi s/(2k_0)$, and vanishes when $c_{\rm WQW} = 2m\pi s/(2k_0)$ ($m \geq 0$ integer), with $\pi s/(2k_0) \approx (0.16\,{\rm nm})s$. This sets constraints on the rise/fall length $\ell = c_{\rm WQW}/s$ of the trapezoidal wiggles.

[1]The dot radius is defined as $r_{\|} = \sqrt{\langle x^2\rangle + \langle y^2\rangle}$, where $\langle x^2\rangle$ and $\langle y^2\rangle$ are the expectation values of $x$ and $y$ in the ground-state.

[2]Here the position $z$ is measured with respect to the middle of the well.

[3]This expression directly follows from the fact that a trapezoidal profile is the convolution of two square profiles.

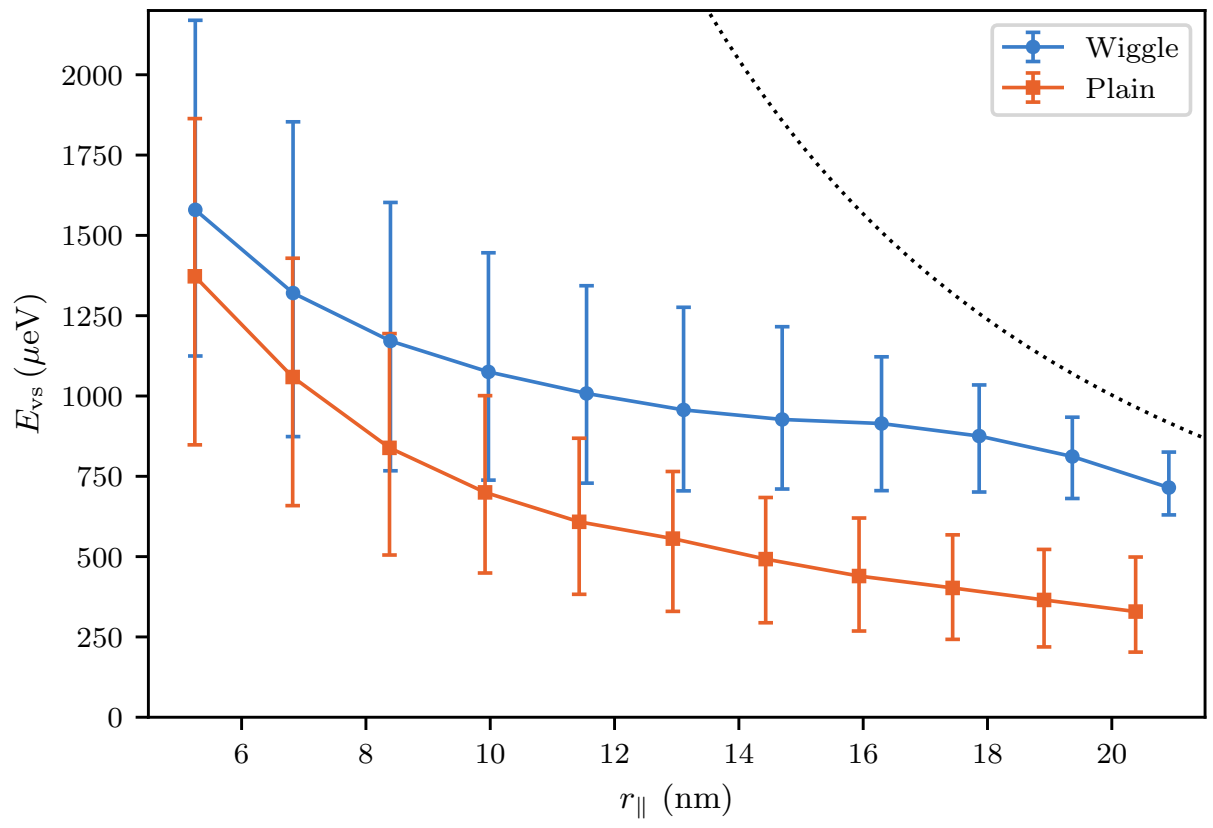


Figure 13: Valley splittings (computed as the difference between the energy of the first two levels) as a function of the dot size $r_\parallel$ in sample B and in the plain quantum well with same average Ge concentration $c_{avg} = 11.3\%$. The median splittings are plotted with error bars giving the inter-quartile range. The black dotted line is the expected energy of the first orbital excitation, $E_{orb} = \hbar^2/(m_t^* r_\parallel^2)$, with $m_t^* = 0.191\, m_0$ the transverse effective mass in silicon. The vertical electric field is $E_z = -5$ mV/nm (2 bands $\mathbf{k}\cdot\mathbf{p}$ calculations [3]).

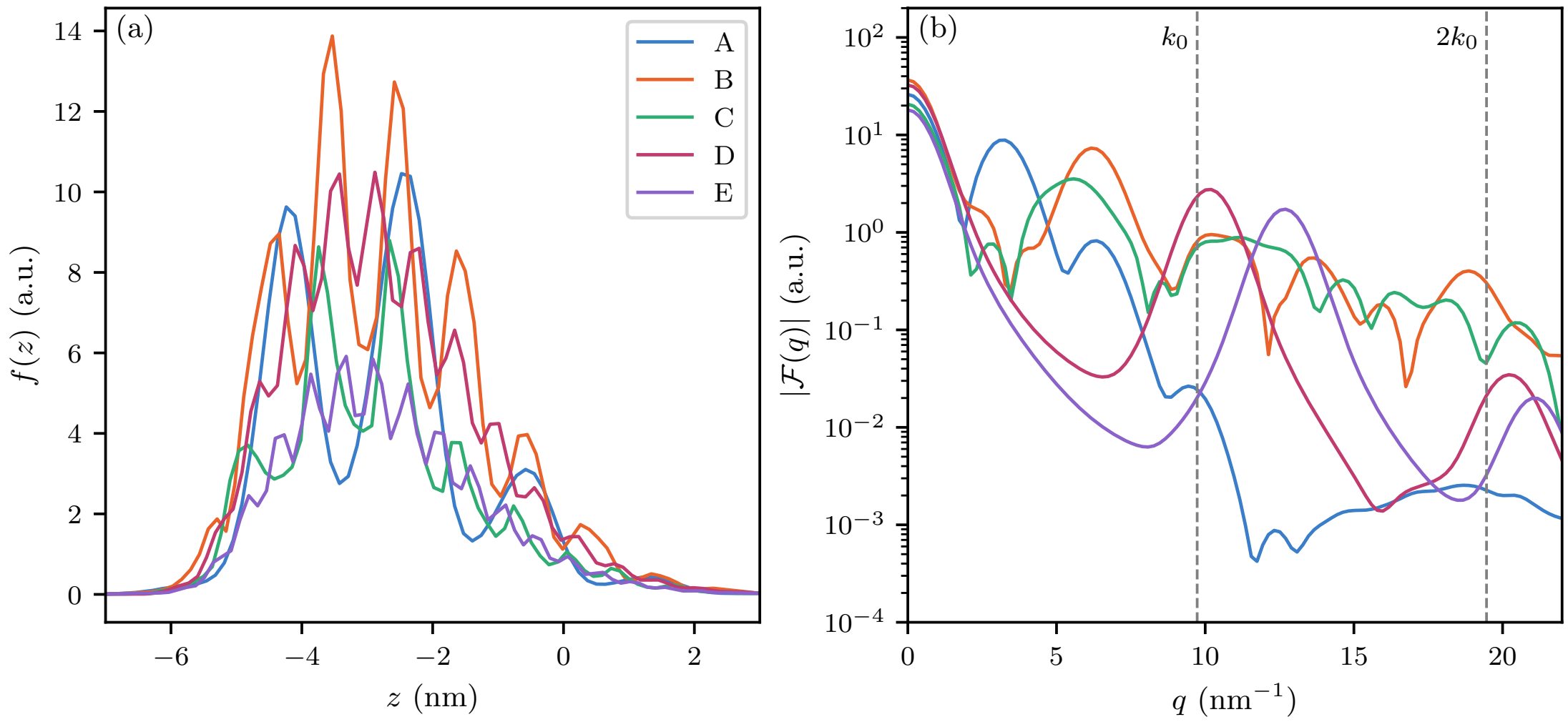


Figure 14: (a) Plots of $f(z) = c_{Ge}(z)|\psi(z)|^2$ for all samples, where $c_{Ge}(z)$ is the Ge concentration profile and $\psi(z)$ is the envelope function in a $L = 80$ monolayers ($\approx 11$ nm) thick quantum well at vertical electric field $E_z = -5$ meV/nm. (b) Fourier transform of $f(z)$ as a function of wave number $q$. The dashed vertical lines mark $k_0 = 9.84\,\text{nm}^{-1}$ and $2k_0 = 19.67\,\text{nm}^{-1}$ (as given by the two bands $\mathbf{k}\cdot\mathbf{p}$ model [3]).

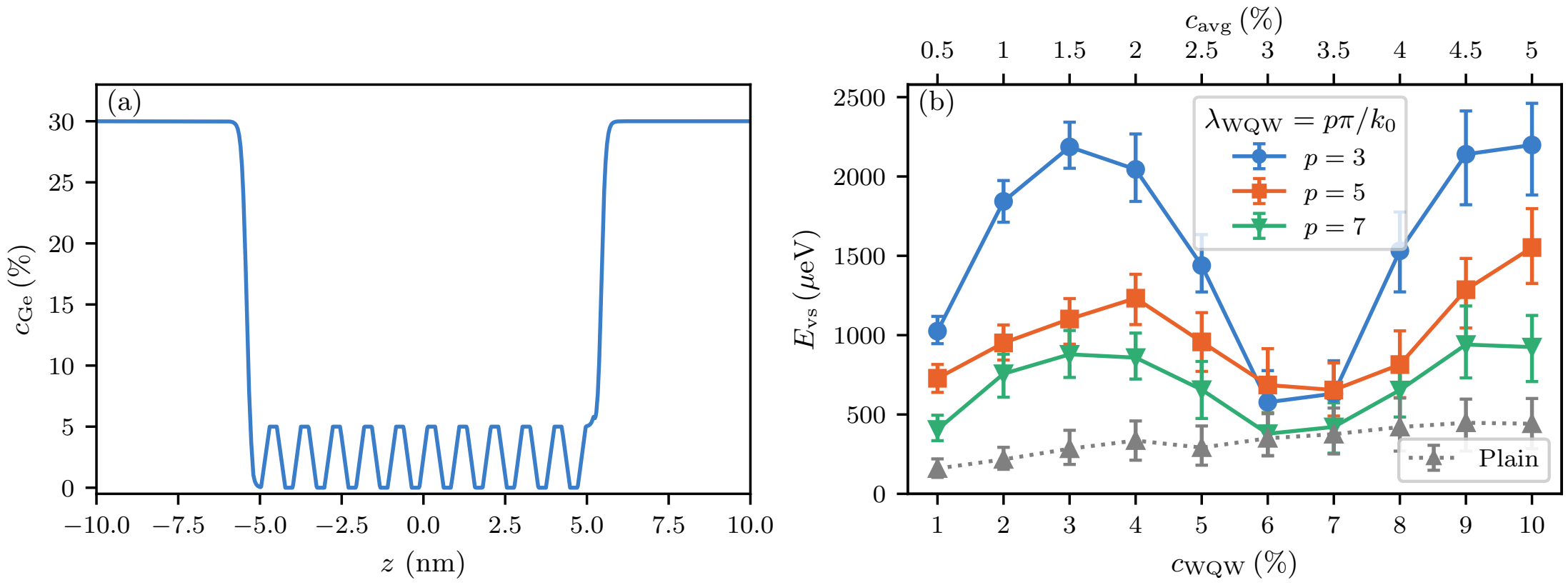


Figure 15: (a) Ge concentration profile in a trapezoidal wiggle well profile with wave length $\lambda_{\mathrm{WQW}} = 3\pi/k_0$, amplitude $c_{\mathrm{WQW}} = 5$ %, and rise/fall slope $s = 20$%/nm. The width of the well is $L = 80$ monolayers ($\approx 11$ nm). (b) Valley splittings as a function of $c_{\mathrm{WQW}}$ in trapezoidal wiggle wells with wave lengths $\lambda_{\mathrm{WQW}} = p\pi/k_0$ ($s = 20$%/nm), and in plain quantum wells with same average Ge concentration $c_{\mathrm{avg}} = c_{\mathrm{WQW}}/2$. The median splittings are plotted with error bars giving the inter-quartile range. The width of the well is $L \approx 11$ nm, the vertical electric field is $E_z = 5$ mV/nm, and the dot radius is $r_{\parallel} = 9.3$ nm (2 bands $\mathbf{k}\cdot\mathbf{p}$ calculations [3]).

To illustrate this behavior, we plot in Fig. 15b the valley splittings computed in such a structure as a function of the Ge concentration $c_{\mathrm{WQW}}$ in the SiGe layers ($s = 20$%/nm as achieved in the experiments). The width of the well is $L = 80$ monolayers ($\approx 11$ nm) and the interfaces with the $\mathrm{Si_{0.7}Ge_{0.3}}$ buffer are smoothed over $w_{\mathrm{int}} = 2$ monolayers [3]. The radius of the dots is $r_{\parallel} = 9.3$ nm (to make sure that the orbital splitting remains larger than the valley splitting). The median valley splitting indeed oscillates with $c_{\mathrm{WQW}}$ and reaches a first maximum at a small $c_{\mathrm{WQW}} \approx 3\%$ (as expected). At that Ge concentration the valley splittings are typically larger than 2 meV. The median valley splitting then decreases down to a minimum at $c_{\mathrm{WQW}} \approx 6.5\%$. There the residual valley splitting is in fact dominated by alloy disorder and is thus comparable to a plain quantum well with same average Ge concentration $c_{\mathrm{avg}} = c_{\mathrm{WQW}}/2$. It then shows an other (equivalent) maximum at $c_{\mathrm{WQW}} \approx 9\%$. Atomistic tight-binding calculations [3, 4] reproduce the same trends.

We plot in Fig. 16 the valley splittings computed in sample B as a function of the vertical electric field $E_z$ (sample B being actually the only one showing a strong dependence on $E_z$). The median valley splitting in the corresponding plain well slightly increases with increasing $|E_z|$ as the electron penetrates farther into the disordered $\mathrm{Si_{0.7}Ge_{0.3}}$ buffer or top barrier (whose interfaces are slightly different, as in the experiments) [2]. At negative electric fields, sample B exhibits robust valley splitting enhancements and most likely operates near the second maximum of Fig. 15 at $c_{\mathrm{WQW}} \approx 9\%$. For positive electric fields, the electron gets localized on the opposite, top interface where the amplitude of the modulations is smaller (see Fig. 2 of the main text) and actually closer to the dip at $c_{\mathrm{WQW}} \approx 6.5\%$. As a consequence, the enhancement of the valley splittings on that interface is much weaker (though still noticeable). Yet sample B was not originally designed as a trapezoidal WQW; we anticipate that a better control of the amplitude and period of the wiggles will allow for robust valley splittings whatever the vertical electric field. In particular, large valley splittings shall be achieved at low wiggle amplitude $c_{\mathrm{WQW}} \approx 3\%$, thus low disorder in the WQW. In the ideal trapezoidal $2k_0/3$ WQW with $s = 20$%/nm and $c_{\mathrm{WQW}} = 3.2\%$ (Fig. 15), the median valley splittings range between 2.19 and 2.24 meV whatever the vertical electric field $|E_z| \leq 5$ mV/nm.

We can, moreover, generally consider trapezoidal wiggle wells with wave length $\lambda_{\mathrm{WQW}} = p\pi/k_0$ ($p = 3, 5, 7$... an odd integer), whose $p^{\mathrm{th}}$ harmonic is resonant with $q = 2k_0$. Its amplitude is

$$|C_p| = \frac{2c_{\mathrm{WQW}}}{p\pi}\left|\mathrm{sinc}\left(\frac{c_{\mathrm{WQW}}k_0}{s}\right)\right| = \frac{2s}{p\pi k_0}\left|\sin\left(\frac{c_{\mathrm{WQW}}k_0}{s}\right)\right|. \tag{5}$$

With respect to the $p = 3$ case, the valley splittings is expected to be scaled down by a factor $3/p$ (the density of wiggles in the well) as long as the wave function probes a reasonable number of those wiggles and does not significantly localize in the thicker Si layers. This is roughly the case for $p = 5$ and $p = 7$, as shown in Fig. 15b. We

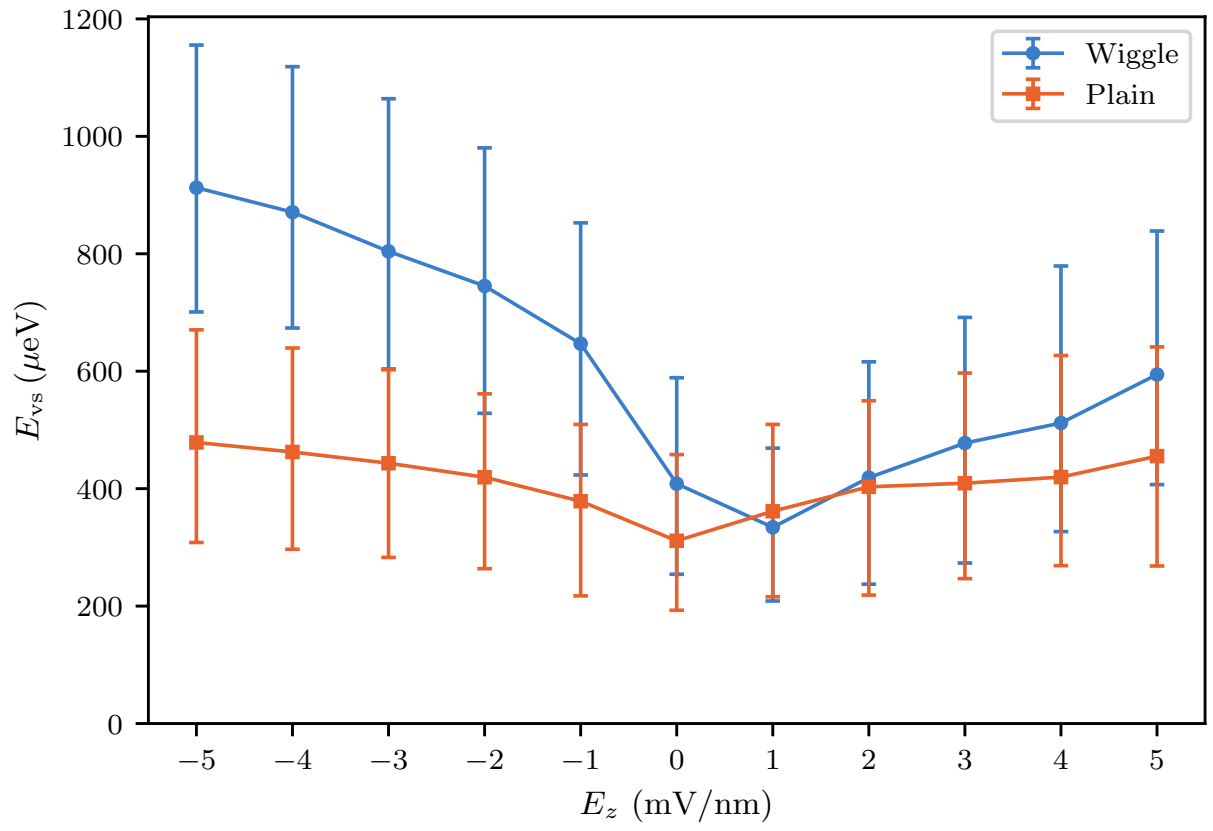


Figure 16: Valley splittings (computed as the difference between the energy of the first two levels) as a function of the vertical electric field $E_z$ in sample B and in the plain quantum well with same average Ge concentration $c_{\rm avg} = 11.3\%$ (dot radius $r_\parallel = 15$ nm). The median splittings are plotted with error bars giving the inter-quartile range (2 bands $\mathbf{k}\cdot\mathbf{p}$ calculations [3]).

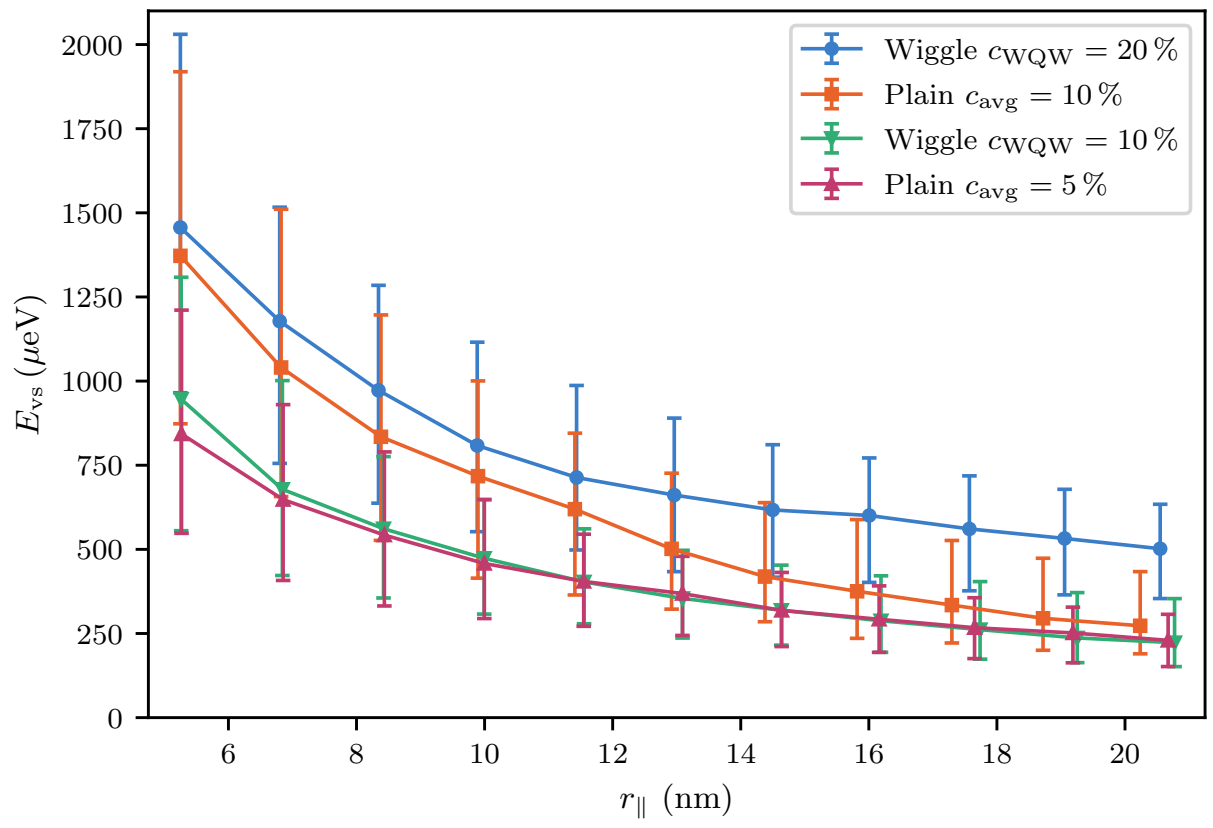


Figure 17: Valley splittings as a function of the dot size $r_\parallel$ in sinusoidal wiggle wells with wave length $\lambda_{\rm WQW} = 2\pi/k_0$ and Ge modulations $c_{\rm WQW} = 10$ or 20%, and in plain quantum wells with same average Ge concentration $c_{\rm avg} = c_{\rm WQW}/2$. The median splittings are plotted with error bars giving the inter-quartile range. The width of the well is $L \approx 11$ nm and the vertical electric field is $E_z = 5$ mV/nm (2 bands $\mathbf{k}\cdot\mathbf{p}$ calculations [3]).

emphasize, though, that the efficiency will decrease (and the variability increase) once $\lambda_{\rm WQW}$ gets comparable to the vertical extent of the wave function (in Fourier space, the nearby harmonics broadened by structural disorder and the envelope function start to overlap and interfere significantly).

We finally plot in Fig. 17 the valley splittings as a function of the dot radius $r_\parallel$ in sinusoidal wiggle wells with wave length $\lambda_{\rm WQW} = 2\pi/k_0$ and different Ge modulations $c_{\rm WQW}$, and in plain quantum wells with same average Ge content $c_{\rm avg} = c_{\rm WQW}/2$. The well width is again $L = 80$ monolayers and the interfaces are smoothed over $w_{\rm int} = 2$ monolayers [3]. The Ge concentration profile $c_{\rm Ge}(z) = c_{\rm WQW}(1 - \cos k_0 z)/2$ is perfectly sinusoidal in the well. At low $c_{\rm WQW} = 10\%$, the effects of the wiggles are superseded by those of alloy disorder (the data being barely distinguishable from $c_{\rm avg} = 5\%$ whatever $r_\parallel < 25$ nm). However, the wiggles deterministically enhance the valley splittings for larger $c_{\rm WQW} = 20\%$, the median being significantly greater than in the plain quantum well with $c_{\rm avg} = 10\%$ and far less dependent on dot size. As a matter of fact, this enhancement results from a non-linear mechanism [5]. Indeed, the envelope function $\psi(z)$ itself gets modulated at wave length $\lambda = 2\pi/k_0$ by the wiggle well potential, so that $f(z) = c_{\rm Ge}(z)|\psi(z)|^2$ has a harmonic at $q = 2k_0$. As these modulations are proportional to $c_{\rm WQW}$ to first order in perturbation, the median valley splitting is expected to increase as $c_{\rm WQW}^2$. We find, practically, a much smoother (but still non-linear) dependence (Fig. 7 of the main text), presumably due to the competition with alloy disorder and to

the close proximity of the orbital excitation at $E_{\mathrm{orb}} = 1.8\,\mathrm{meV}$.



\end{document}